\documentclass{ar-1col-S2O}
\usepackage[numbers,sort&compress]{natbib}
\usepackage{url}
\jname{Annu. Rev. Nucl. Part. Sci.}
\jvol{75}
\jyear{2025}
\doi{10.1146/annurev-nucl-102122-023242}

\usepackage{amsmath}
\usepackage{amsfonts}
\usepackage{amssymb}
\usepackage{graphicx}
\usepackage{bm}
\usepackage{subfigure}
\usepackage[hyperindex]{hyperref}
\usepackage{bigdelim}
\usepackage{booktabs}
\usepackage{dcolumn}
\usepackage{multirow}
\usepackage{cancel}
\usepackage{stackrel}
\usepackage{paralist}
\usepackage{xspace}
\usepackage{slashed}
\usepackage{cancel}
\usepackage{enumerate}
\usepackage{float}
\usepackage{fullpage}
\usepackage{ulem}
\usepackage{footmisc}

\def\cevns{\text{CE$\nu$NS}\xspace}
\def\cevas{\text{CE$\nu$AS}\xspace}
\def\eves{\text{E$\nu$ES}\xspace}
\newcommand{\vet}[1]{\ensuremath{\hskip-1pt\vec{\hskip1pt#1}}}

\newcommand{\red}[1]{#1}

\begin{document}

\markboth{Giunti et al.}{Neutrino Electromagnetic Properties}

\title{Neutrino Electromagnetic Properties}

\author{Carlo Giunti,$^1$ Konstantin Kouzakov,$^2$ Yu-Feng~Li,$^3$ and Alexander Studenikin$^4$
\affil{$^1$Istituto Nazionale di Fisica Nucleare (INFN), Sezione di Torino, Via~P.~Giuria~1, 10125 Torino, Italy;
email:~carlo.giunti@to.infn.it}
\affil{$^2$Faculty of Physics, Lomonosov Moscow State University, Leninskie Gory, Moscow 119991, Russia;
email:~kouzakov@gmail.com}
\affil{$^3$Institute of High Energy Physics, Chinese Academy of Sciences, Beijing 100049, China;
School of Physical Sciences, University of Chinese Academy of Sciences, Beijing 100049, China;
email:~liyufeng@ihep.ac.cn}
\affil{$^4$Faculty of Physics, Lomonosov Moscow State University, Leninskie Gory, Moscow 119991, Russia;
email:~studenik@srd.sinp.msu.ru}}

\begin{abstract}
Neutrinos are neutral in the Standard Model,
but they have tiny charge radii
generated by radiative corrections.
In theories Beyond the Standard Model,
neutrinos can also have magnetic and electric moments
and
small electric charges (millicharges).
We review the general theory
of neutrino electromagnetic form factors,
which reduce,
for ultrarelativistic neutrinos and small momentum transfers,
to the
neutrino charges, effective charge radii, and effective magnetic moments.
We discuss the phenomenology of these electromagnetic neutrino properties
and we review the existing experimental bounds.
We briefly review also
the electromagnetic processes of astrophysical neutrinos
and the
neutrino magnetic moment portal
in the presence of sterile neutrinos.
\end{abstract}


\maketitle

\tableofcontents

\section{Introduction}
\label{sec:intro}

Neutrinos are the most elusive among all the known elementary particles
because their interactions with matter particles (quarks and electrons)
are feeble.
For this reason, some of their fundamental properties are not known:
most important are the absolute values of the neutrino masses and if
neutrinos are Dirac fermions, endowed with corresponding antiparticles as quarks and charged leptons,
or Majorana fermions,
which coincide with their antiparticles.

Most of the experimentally known properties of neutrinos are in agreement with the
Standard Model (SM) of electroweak interactions.
The most important exception is the fact that neutrinos are massive,
contrary to the SM prediction,
as was discovered in oscillation experiments
(see, e.g., Refs.\citenum{ParticleDataGroup:2024cfk,Giunti:2007ry,Bilenky:2018hbz}).
Therefore,
the SM must be extended to accommodate neutrino masses.
There are also other reasons to extend the SM,
as the necessity to explain the
plausible existence of Dark Matter and Dark Energy and
the matter-antimatter asymmetry in the Universe.
There are many models which perform these tasks and implement the
so-called new physics Beyond the Standard Model (BSM)~\cite{Coloma:2022dng}.

In the SM neutrinos are electrically neutral
and do not interact directly
(i.e. at the tree level) with electromagnetic fields.
Therefore,
the only electromagnetic properties of neutrinos in the SM
are the tiny charge radii generated by radiative corrections.
In BSM models neutrinos can have also
magnetic and electric moments,
which are often proportional to the neutrino masses,
and tiny electric charges.

The neutrino magnetic moment $\mu_\nu$ was introduced by Wolfgang Pauli in his famous letter \cite{Pauli:1930pc} addressed to "Dear Radioactive Ladies and Gentlemen" of the  T\"ubingen meeting of the German Physical Society on December 1930.
The impact of neutrinos with nonzero magnetic moment propagating in matter was studied
for the first time in 1932 by Carlson and Oppenheimer~\cite{Carlson:1932rk}.
The cross section of magnetic moment interaction of
a neutrino with an electron was calculated by Bethe in
1935~\cite{Bethe:1935cp}.
Before the first observation of neutrinos by Cowan and Reines in 1956~\cite{Cowan:1956rrn}
through the inverse beta decay (IBD) process
$\nu_{e} + p \to n + e^{+}$,
whose cross section was estimated to be extremely small by
Bethe and Peierls~\cite{Bethe:1934qn},
the magnetic moment interaction was considered as an alternative
process for detecting neutrinos if the neutrino magnetic moment is large enough.
The first attempt to detect neutrinos through its magnetic moment interaction
was done in 1935 by Nahmias~\cite{Nahmias:1935yih}
with two Geiger-M\"uller counters placed near a radioactive source.
The neutrinos were not detected and Nahmias established
the upper bound\footnote{$\mu_{\text{B}} \equiv e / 2 m_{e}$
is the Bohr magneton, where $e$ is the elementary charge and $m_{e}$ is the electron mass.}
$\mu_\nu < 2 \times 10^{-4} \, \mu_{\text{B}}$.
Before their first observation of neutrinos in 1956~\cite{Cowan:1956rrn},
in 1954 Cowan and Reines
attempted to measure the neutrino magnetic moment
with a liquid scintillation detector exposed
to the electron antineutrino flux of the
Savannah River reactor.
They obtained the upper limit
$\mu_\nu < 10^{-7} \mu_{\text{B}}$~\cite{Cowan:1954pq}.
Later, in 1957,
they obtained
$\mu_\nu < 10^{-9} \mu_{\text{B}}$~\cite{Cowan:1957pp}
with a larger and better-shielded detector.
Since then,
the electromagnetic properties and interactions of neutrinos
have been studied in many theoretical and experimental works
(see, e.g., Refs.\citenum{Bernstein:1963qh,Raffelt:1996wa,Raffelt:1999tx,Giunti:2014ixa,BahaBalantekin:2018ppj,Das:2020egb,Studenikin:2022rhv,Abdullah:2022zue}
and references therein).

In this review,
first we briefly review in Section~\ref{sec:mixing}
the phenomenology of neutrino mixing and oscillations,
which is useful for the following discussion of neutrino electromagnetic properties.
In Section~\ref{sec:properties}
we review the general theory of the
electromagnetic properties of massive Dirac and Majorana neutrinos.
In Sections~\ref{sec:MM}, \ref{sec:CR}, and \ref{sec:EC}
we review the theory of
neutrino magnetic moments,
charge radii,
and
electric charges (sometimes called ``millicharges'')
and we present the current bounds
obtained directly from experiments and
from phenomenological analyses of experimental data.
In Section~\ref{sec:Astrophysical}
we review the phenomenology
of the most important electromagnetic processes
of astrophysical neutrinos.
In Section~\ref{sec:portal}
we review the so-called
``neutrino magnetic moment portal''
or
``neutrino dipole moment portal'',
which is currently considered an interesting
theoretical possibility.
Finally,
in Section~\ref{sec:future}
we conclude and briefly indicate
the future experiments which can
improve the search of neutrino electromagnetic interactions.

\section{Neutrino mixing and oscillations}
\label{sec:mixing}

We know that there are three active\footnote{Active neutrinos interact with SM weak interactions, sterile neutrinos do not.}
neutrino flavors:
$\nu_{e}$,
$\nu_{\mu}$, and
$\nu_{\tau}$,
which produce the corresponding charged leptons
$e$,
$\mu$, and
$\tau$
in charged-current weak interactions.
In the minimal framework of three-neutrino mixing
(see, e.g., Refs.\citenum{ParticleDataGroup:2024cfk,Giunti:2007ry,Bilenky:2018hbz}),
the three flavor neutrinos are superpositions of three massive neutrinos
$\nu_{1}$,
$\nu_{2}$, and
$\nu_{3}$,
given by\footnote{We use the index $\ell = e, \mu, \tau$
to denote the active flavor neutrinos
$\nu_{e}$,
$\nu_{\mu}$,
$\nu_{\tau}$.}
\begin{equation}
| \nu_{\ell} \rangle
=
\sum_{k=1}^{3}
U_{\ell k}^{*} \, | \nu_{k} \rangle
,
\quad
\text{for}
\quad
\ell = e, \mu, \tau
,
\label{flav}
\end{equation}
where $U$ is the $3\times3$ unitary mixing matrix.
Neutrino oscillations are flavor transitions which depend on
the elements of the mixing matrix $U$
and on the squared-mass differences
$\Delta{m}^{2}_{kj} \equiv m_{k}^2 - m_{j}^2$.
The massive neutrinos can be either of Dirac or Majorana type.
In the Dirac case,
there are distinct neutrino and antineutrino states,
whereas in the Majorana case neutrino and antineutrino coincide.
The helicity of active Dirac neutrinos is left-handed 
and that of active Dirac antineutrinos is right-handed.
Active Majorana neutrinos can be either left-handed or right-handed.
Hence, it is convenient to call ``neutrino''
both
a left-handed Dirac neutrino
or
a left-handed Majorana neutrino
and to call ``antineutrino''
both
a right-handed Dirac antineutrino
or
a right-handed Majorana neutrino.
Antineutrinos obey a mixing relation similar to~\ref{flav},
with $U^{*} \to U$.
The three neutrino masses are constrained below about 1 eV
by kinematical measurements~\cite{ParticleDataGroup:2024cfk}.

The $3\times3$ unitary mixing matrix can be parameterized in terms of
three mixing angles,
one Dirac CP-violating phase,
and
two Majorana CP-violating phases
in the case of Majorana neutrinos.
Neglecting the Majorana phases,
which are completely unknown and
do not have any effect on neutrino oscillations,
except in the case of neutrino spin oscillations
in the presence of magnetic fields and matter
(see Section~\ref{sec:spin_oscillations}),
the standard parameterization of the mixing matrix is
\begin{equation}
U =
\begin{pmatrix}
c_{12} c_{13} & s_{12} c_{13} & s_{13} e^{-i\delta_{13}}
\\
- s_{12} c_{23} - c_{12} s_{23} s_{13} e^{i\delta_{13}} & c_{12} c_{23} -
s_{12} s_{23} s_{13} e^{i\delta_{13}} & s_{23} c_{13}
\\
s_{12} s_{23} - c_{12} c_{23} s_{13} e^{i\delta_{13}} & - c_{12} s_{23} -
s_{12} c_{23} s_{13} e^{i\delta_{13}} & c_{23} c_{13}
\end{pmatrix}
,
\label{mixmat}
\end{equation}
where $ c_{ab} \equiv \cos\vartheta_{ab} $ and $ s_{ab} \equiv
\sin\vartheta_{ab} $.
The values of the three mixing angles
$\vartheta_{12}$,
$\vartheta_{23}$, and
$\vartheta_{13}$
are constrained between $0$ and $\pi/2$,
whereas the value of the Dirac CP-violating phase $\delta_{13}$
can be between $0$ and $2\pi$.
This parameterization is convenient because
it has the simplest form of the first line
concerning electron neutrinos,
which are produced and/or detected in many experiments.
It is also convenient because the CP-violating phase $\delta_{13}$
is associated with the smallest mixing angle $\vartheta_{13}$
and it is clear that the amount of CP violation is proportional to
$\sin\vartheta_{13}$.

Neutrino oscillations have been discovered in 1998 in the Super-Kamiokande
atmospheric neutrino experiment~\cite{Super-Kamiokande:1998kpq}
and later confirmed in several long-baseline accelerator and reactor experiments
and in solar neutrino experiments
(see, e.g., Refs.\citenum{ParticleDataGroup:2024cfk,Giunti:2007ry,Bilenky:2018hbz}).
The analysis of the data of these experiments
established that
$\Delta{m}^2_{21} \ll |\Delta{m}^2_{31}|$,
where $\Delta{m}^2_{21}>0$ by convention ($m_{2}>m_{1}$),
but the sign of $\Delta{m}^2_{31}$ is not known.
Hence,
there are two possible orderings of the neutrino masses:
the normal ordering
(NO)
with
$m_{1}<m_{2}<m_{3}$
and
$\Delta{m}^{2}_{31}, \, \Delta{m}^{2}_{32} > 0$;
the inverted ordering
(IO)
with
$m_{3}<m_{1}<m_{2}$
and
$\Delta{m}^{2}_{31}, \, \Delta{m}^{2}_{32} < 0$.
The best-fit values of the oscillation parameters
obtained from a global fit
of neutrino oscillation data~\cite{deSalas:2020pgw}
(see also the similar results obtained in Refs.\citenum{Esteban:2020cvm,Capozzi:2021fjo})
are
\begin{equation}
\renewcommand{\arraystretch}{1.45}
\setlength{\arraycolsep}{1cm}
\begin{array}{ll} \displaystyle
\Delta{m}^2_{21} = 7.50^{+0.22}_{-0.20} \times 10^{-5} \, \text{eV}^2
,
& \displaystyle
\sin^2\vartheta_{12} = 0.318 \pm 0.016
,
\\ \displaystyle
\Delta{m}^2_{31}|_{\text{NO}} = 2.55^{+0.02}_{-0.03} \times 10^{-3} \, \text{eV}^2
,
& \displaystyle
\Delta{m}^2_{31}|_{\text{IO}} = - 2.45^{+0.02}_{-0.03} \times 10^{-3} \, \text{eV}^2
,
\\ \displaystyle
\left. \sin^2\vartheta_{23} \right|_{\text{NO}} = 0.574 \pm 0.014
,
& \displaystyle
\left. \sin^2\vartheta_{23} \right|_{\text{IO}} = 0.578^{+0.010}_{-0.017}
,
\\ \displaystyle
\left. \sin^2\vartheta_{13} \right|_{\text{NO}} = 0.02200^{+0.00069}_{-0.00062}
,
& \displaystyle
\left. \sin^2\vartheta_{13} \right|_{\text{IO}} = 0.02225^{+0.00064}_{-0.00070}
\\ \displaystyle
\left. \delta_{13} \right|_{\text{NO}} = 1.08^{+0.13}_{-0.12} \pi
,
& \displaystyle
\left. \delta_{13} \right|_{\text{IO}} = 1.58^{+0.15}_{-0.16} \pi
,
\end{array}
\label{mixbf}
\end{equation}
The best-fit values of
$|\Delta{m}^2_{31}|$,
$\sin^2\vartheta_{23}$,
$\sin^2\vartheta_{13}$, and
$\delta_{13}$
are slightly different in the NO and IO.
There is a $2.5\sigma$
preference for NO,
which however is not considered as the definitive establishment of NO.
The issue is under investigation in new experiments
(JUNO~\cite{JUNO:2021vlw},
DUNE~\cite{DUNE:2020jqi},
Hyper-Kamiokande~\cite{Hyper-Kamiokande:2022smq},
ESSnuSB~\cite{ESSnuSB:2021azq}).
Although the mixing angle $\vartheta_{23}$
seems to be relatively well-determined from the $1\sigma$
uncertainties in Eq.\ref{mixbf},
it is the less known mixing angle,
with $3\sigma$ ranges
$ \left. \sin^2\vartheta_{23} \right|_{\text{NO}} \in (0.434, 0.610)$
and
$ \left. \sin^2\vartheta_{23} \right|_{\text{IO}} \in (0.433, 0.608)$.
Therefore,
it is not known if $\vartheta_{23}$ is smaller, larger or equal to
the maximal mixing value of $\pi/4$ (i.e. $\sin^2\vartheta_{23} = 0.5$).

\section{Electromagnetic properties of massive Dirac and Majorana neutrinos}
\label{sec:properties}

Neutrinos are generally believed to be neutral particles which do not have electromagnetic interactions.
This belief stems from the fact that in the SM
neutrinos are exactly neutral
and there are no electromagnetic interactions at the tree-level.
However,
even in the SM neutrinos
are predicted to have small electromagnetic interactions
due to the so-called ``charge radius''
\cite{Bernabeu:2000hf,Bernabeu:2002nw,Bernabeu:2002pd,Erler:2013xha}
which is generated by radiative corrections.
Moreover, in BSM theories,
neutrinos can have additional electromagnetic interactions.

\begin{figure}
\includegraphics[height=3cm]{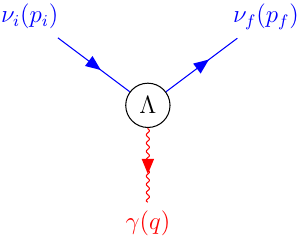}
\caption{Effective coupling of massive neutrinos with a photon.
The initial ($\nu_{i}$) and final ($\nu_{f}$) massive neutrinos
can be different.}
\label{fig:EMvertex}
\end{figure}

The electromagnetic properties of neutrinos
are derived from a general description of
neutrino electromagnetic interactions
in the one-photon approximation
through the effective interaction Hamiltonian
\begin{equation}
\mathcal{H}_{\text{em}}^{(\nu)}(x)
=
j^{\mu}_{(\nu)}(x) A_{\mu}(x)
,
\label{eq:Hamiltonian}
\end{equation}
where $A_{\mu}(x)$ is the electromagnetic field
and
\begin{equation}
j^{\mu}_{(\nu)}(x)
=
\sum_{k,j}
\overline{\nu_{k}}(x) \widehat\Lambda^{\mu}_{kj} \nu_{j}(x)
\label{eq:current}
\end{equation}
is the effective electromagnetic current four-vector.
$\widehat\Lambda^{\mu}_{kj}$ is
the effective electromagnetic operator in spinor space.
The sum in Eq.\ref{eq:current}
is over the indices $k,j=1,2,3,\ldots$
of the massive neutrinos,
which are three in the case of standard three-neutrino mixing,
but can be more if additional sterile neutrinos exist
\cite{Gariazzo:2015rra,Drewes:2016upu,Boyarsky:2018tvu,Giunti:2019aiy,Boser:2019rta,Dasgupta:2021ies}.
The effective electromagnetic current $j^{\mu}_{(\nu)}(x)$
acts between neutrino massive states in the calculation of
electromagnetic interactions represented by the Feynman diagram in
Fig.\ref{fig:EMvertex}:
\begin{equation}
\langle \nu_{f}(p_{f}) |
j^{\mu}_{(\nu)}(x)
| \nu_{i}(p_{i}) \rangle
=
e^{i(p_{f}-p_{i})x}
\,
\overline{u_{f}}(p_{f})
\Lambda^{\mu}_{fi}(q)
u_{i}(p_{i})
,
\label{eq:matrixelement}
\end{equation}
where the vertex function
$\Lambda^{\mu}_{fi}(q)$
is a $4\times4$ matrix in spinor space
which depends on the four-momentum transfer $q=p_{i}-p_{f}$
(the four-momentum of the photon in Fig.\ref{fig:EMvertex}).
The vertex function can be decomposed
in a sum of linearly independent products
of Dirac $\gamma$ matrices.
Then,
from the hermiticity of the interaction Hamiltonian \ref{eq:Hamiltonian}
and gauge invariance,
it can be reduced to
\begin{equation}
\Lambda^{\mu}_{fi}(q)
=
\left( \gamma^{\mu} - q^{\mu} \slashed{q}/q^{2} \right)
\left[
F^{Q}_{fi}(q^{2})
+
F^{A}_{fi}(q^{2}) q^{2} \gamma_{5}
\right]
-
i \sigma^{\mu\nu} q_{\nu}
\left[
F^{M}_{fi}(q^{2})
+ i
F^{E}_{fi}(q^{2}) \gamma_{5}
\right]
,
\label{eq:Lambda}
\end{equation}
with the four
charge ($F^{Q}_{fi}(q^{2})$),
anapole ($F^{A}_{fi}(q^{2})$),
magnetic ($F^{M}_{fi}(q^{2})$), and
electric ($F^{E}_{fi}(q^{2})$)
form factors,
having the following main characteristics:
\begin{enumerate}

\renewcommand{\labelenumi}{(\theenumi)}
\renewcommand{\theenumi}{\alph{enumi}}

\item
\label{propa}
The matrices of the form factors
$F^{X}$,
for $X=Q,A,M,E$,
are Hermitian:
\begin{equation}
F^{X} = F^{X\dagger}
\quad
\Longleftrightarrow
\quad
F^{X}_{fi}=(F^{X}_{if})^{*}
.
\label{hermitian}
\end{equation}
Therefore,
the diagonal form factors $F^{X}_{ii}$
are real.

\item
\label{propb}
For antineutrinos
\begin{align}
\overline{F}^{X}
=
- (F^{X})^{T}
=
- (F^{X})^{*}
\null & \null
\quad
\Longleftrightarrow
\quad
\overline{F}^{X}_{fi} = - F^{X}_{if} = - (F^{X}_{fi})^{*}
\quad
\text{for}
\quad
X=Q,M,E
,
\label{anti1}
\\
\overline{F}^{A}
=
(\overline{F}^{A})^{T}
=
(\overline{F}^{A})^{*}
\null & \null
\quad
\Longleftrightarrow
\quad
\overline{F}^{A}_{fi} = F^{A}_{if} = (F^{A}_{fi})^{*}
.
\label{anti2}
\end{align}
Therefore, the real diagonal charge, magnetic and electric form factors
of antineutrinos
have the same size and opposite signs
of those of neutrinos
($ \overline{F}^{Q}_{ii} = - F^{Q}_{ii} $,
$ \overline{F}^{M}_{ii} = - F^{M}_{ii} $, and
$ \overline{F}^{E}_{ii} = - F^{E}_{ii} $),
as expected.
The real diagonal anapole form factors are the same for neutrinos and antineutrinos.

\item
\label{propc}
For Majorana neutrinos the neutrino-antineutrino equality
implies that the matrices of the charge, magnetic and electric form factors
are antisymmetric
and the matrix of the anapole form factor is symmetric:
\begin{align}
F^{X} = - (F^{X})^{T} = - (F^{X})^{*}
\null & \null
\quad
\Longleftrightarrow
\quad
F^{X}_{fi} = - F^{X}_{if} = - (F^{X}_{fi})^{*}
\quad
\text{for}
\quad
X=Q,M,E
,
\label{ffmaj1}
\\
F^{A} = (F^{A})^{T} = (F^{A})^{*}
\null & \null
\quad
\Longleftrightarrow
\quad
F^{A}_{fi} = F^{A}_{if} = (F^{A}_{fi})^{*}
.
\label{ffmaj2}
\end{align}
Therefore,
Majorana neutrinos
do not have diagonal charge, magnetic, and electric form factors
($ F^{X}_{ii} = 0 $
for
$ X=Q,M,E $),
but can have off-diagonal
imaginary charge, magnetic, and electric form factors
$ F^{Q}_{fi} $,
$ F^{M}_{fi} $, and
$ F^{E}_{fi} $
for $ f \neq i $.
On the other hand,
Majorana neutrinos can have real diagonal and off-diagonal anapole form factors.

\end{enumerate}

For the coupling with a real photon ($q^{2}=0$)
the four form factors reduce, respectively, to the
electric charge $Q_{fi}$,
anapole $A_{fi}$,
magnetic moment $\mu_{fi}$, and
electric moment $\varepsilon_{fi}$
matrices of massive neutrinos.
Although this limit of a constant value is strictly true only for $q^{2}=0$,
it has been so far adopted also for electromagnetic neutrino interactions
mediated by virtual photons ($q^{2} \neq 0$),
because there is no theoretical prediction for the $q^{2}$ dependence of the form factors.
Only the first-order $q^{2}$ dependence of the charge form factors
$F^{Q}_{fi}(q^{2})$
corresponding to the massive neutrino charge radii\footnote{Strictly speaking
$\langle r_{fi}^2 \rangle$
is the square of a charge radius,
but for simplicity it is commonly called charge radius.
The interpretation of the first-order $q^{2}$ dependence of the charge form factors
in terms of charge radii follow from the interpretation of a charge form factor
as the Fourier transform of the charge distribution
(see, e.g, Ref.\citenum{Giunti:2014ixa}).}
$\langle r_{fi}^2 \rangle$
is considered:
\begin{equation}
F^{Q}_{fi}(q^{2})
\simeq
F^{Q}_{fi}(0)
+
q^{2}
\left.
\dfrac{ d F^{Q}_{fi}(q^{2}) }{ d q^{2} }
\right|_{q^{2}=0}
=
Q_{fi}
+
\dfrac{ q^{2} }{ 6 } \, \langle r_{fi}^2 \rangle
.
\label{eq:FQ}
\end{equation}
The reason is that
the SM predicts the existence of small neutrino charge radii
(see Section~\ref{sec:CR})
which must be taken into account.
Moreover, if neutrinos are neutral ($Q_{fi}=0$),
as predicted by the SM and many BSM theories,
the leading contributions of the charge form factors
to neutrino electromagnetic interactions at low $q^{2}$
are given by the charge radii.

Since in Eq.\ref{eq:matrixelement} the vertex function
is inserted among the spinors of ultrarelativistic left-handed neutrinos,
the effect of $\gamma_{5}$ is a simple change of sign of the anapole and electric moment terms.
Therefore,
at low $q^{2}$ the vertex function can be reduced to the effective form
\begin{equation}
\Lambda^{\mu}_{fi}(q)
=
\left( \gamma^{\mu} - q^{\mu} \slashed{q}/q^{2} \right)
\left[
Q_{fi}
+
q^{2} \langle r_{fi}^2 \rangle / 6
-
q^{2} A_{fi}
\right]
-
i \sigma^{\mu\nu} q_{\nu}
\left[
\mu_{fi}
- i
\varepsilon_{fi}
\right]
.
\label{eq:Lambda1}
\end{equation}
The term $q_{\mu} \slashed{q}/q^{2}$
can be omitted,
because it disappears for the coupling with a real photon,
as in radiative neutrino decay
$ \nu_{i} \to \nu_{f} + \gamma $,
where it is multiplied to the transverse photon polarization vector,
and in scattering processes
where it is multiplied to a conserved current.
Hence,
we consider the effective vertex function
\begin{equation}
\Lambda^{\mu}_{fi}(q)
=
\gamma^{\mu}
\left[
Q_{fi}
+
q^{2} \langle r_{fi}^2 \rangle^{\text{eff}} / 6
\right]
-
i \sigma^{\mu\nu} q_{\nu}
\mu_{fi}^{\text{eff}}
,
\label{eq:Lambda2}
\end{equation}
where now
$\langle r_{fi}^2 \rangle^{\text{eff}}$
are effective charge radii which include possible anapole contributions
and
$\mu_{fi}^{\text{eff}}$
are effective magnetic moments which include possible electric moment
contributions.
Note that,
from item (\ref{propc}) above
Majorana neutrinos can have diagonal effective charge radii
given by the anapole contributions.
However,
for simplicity in the following we neglect the anapole
and electric moment contributions and we drop the ``eff'' superscript.

An important property of the effective vertex function is that
the charge and charge radius terms conserve the helicity of ultrarelativistic neutrinos,
as the SM weak interaction,
whereas the magnetic moment term flips the helicity.
Therefore,
the contributions of the charge and charge radius to the cross section of any process is coherent
with the SM contribution and must be added to the amplitude of the process.
On the other hand the contribution of the magnetic moment is incoherent
with the SM contribution and must be added to the cross section of the process.

So far we have considered the neutrino electromagnetic form factors
in the mass basis,
where they are well-defined because massive neutrinos have definite four-momenta
which are required to write Eq.\ref{eq:matrixelement}.
Since the flavor neutrinos defined in Eq.\ref{flav}
are superpositions of different massive neutrinos with different masses,
they do not have definite four-momenta.
However, if we neglect the small differences of the neutrino masses in the
electromagnetic interactions,
we can define the electromagnetic form factors in the flavor basis through the matrix element
\begin{equation}
\langle \nu_{\ell'}(p_{f}) |
j^{\mu}_{(\nu)}(x)
| \nu_{\ell}(p_{i}) \rangle
=
\sum_{j,k}
U_{\ell'j}
\langle \nu_{j}(p_{f}) |
j^{\mu}_{(\nu)}(x)
| \nu_{k}(p_{i}) \rangle
U_{\ell k}^{*}
.
\label{eq:flavme}
\end{equation}
Then, from Eqs.\ref{eq:matrixelement} and~\ref{eq:Lambda}
we obtain, for $X=Q,A,M,E$,
\begin{equation}
\Lambda^{\mu}_{\ell'\ell}
=
\sum_{j,k}
U_{\ell'j}
\Lambda^{\mu}_{jk}
U_{\ell k}^{*}
\quad
\Rightarrow
\quad
F^{X}_{\ell'\ell}
=
\sum_{j,k}
U_{\ell'j}
F^{X}_{jk}
U_{\ell k}^{*}
.
\label{eq:flavff}
\end{equation}
\red{Since the transformation from the mass basis to the flavor basis is unitary,
the hermiticity of the electromagnetic form factors
in the mass basis is preserved in the flavor basis.
The form factors of flavor antineutrinos are given by
the unitary transformation~\ref{eq:flavff}
with $U \to U^{*}$.
Therefore,
we have
\begin{equation}
\overline{F}^{X}_{\ell\ell'} = - F^{X}_{\ell'\ell} = - (F^{X}_{\ell\ell'})^{*}
\quad
\text{for}
\quad
X=Q,M,E,
\quad
\text{and}
\quad
\overline{F}^{A}_{\ell\ell'} = F^{A}_{\ell'\ell} = (F^{A}_{\ell\ell'})^{*}
,
\label{eq:antiflavff}
\end{equation}
as in the mass basis (Eqs.\ref{anti1} and~\ref{anti2}).
In particular, the real diagonal charge, magnetic and electric form factors
of flavor antineutrinos have the same size and opposite signs
of those of flavor neutrinos, as expected.
For Majorana neutrinos the antisymmetry
in Eq.\ref{ffmaj1} of the charge, magnetic and electric form factors
in the mass basis is not preserved
in the flavor basis by the unitary transformation~\ref{eq:flavff}
if the mixing matrix $U$ is not real
(i.e. if there are Dirac or Majorana CP-violating phases).
Therefore,
in general also Majorana neutrinos can have real diagonal charge, magnetic and electric form factors in the flavor basis.
This is not in contradiction with Eq.\ref{eq:antiflavff}
because the Majorana neutrino-antineutrino equality
is not valid in the flavor basis if the mixing matrix $U$ is not real.
}

\section{Neutrino magnetic moment}
\label{sec:MM}

\begin{figure}
\centering
\includegraphics[width=0.9\textwidth]{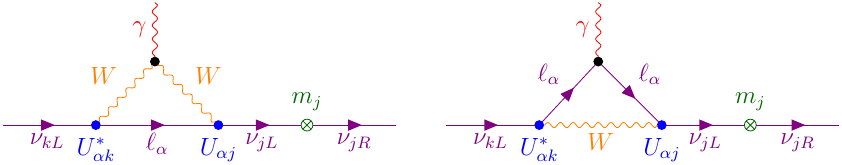}
\caption{Feynman diagrams illustrating the proportionality of the
magnetic moments to the neutrino masses,
which are required to flip the neutrino chirality
through the mass terms
$ m_{j} \overline{\nu_{jR}} \nu_{jL} $.}
\label{fig:chiralityflip}
\end{figure}

The magnetic moments
are the most studied neutrino electromagnetic properties,
because they can be generated in many models with massive neutrinos.
The magnetic moments $\mu_{kj}^{\text{D}}$
of massive Dirac neutrinos are generated
by the effective dimension-five interaction Lagrangian
\begin{equation}
\mathcal{L}_{\text{mag}}^{\text{D}}
=
- \dfrac{1}{2}
\sum_{k,j}
\overline{\nu_{k}}
\,
\sigma^{\alpha\beta}
\mu_{kj}^{\text{D}}
\nu_{j}
\,
F_{\alpha\beta}
=
- \dfrac{1}{2}
\sum_{k,j}
\overline{\nu_{kL}}
\,
\sigma^{\alpha\beta}
\mu_{kj}^{\text{D}}
\nu_{jR}
\,
F_{\alpha\beta}
+
\text{h.c.}
,
\label{dim5lagD}
\end{equation}
where
$\nu_{kL}$ and $\nu_{kR}$
are the left-handed and right-handed components
of the Dirac neutrino field
$\nu_{k}=\nu_{kL}+\nu_{kR}$,
and
$F_{\alpha\beta} = \partial_{\alpha} A_{\beta} - \partial_{\beta} A_{\alpha}$
is the electromagnetic tensor.
In the simplest extension of the SM with three massive Dirac neutrinos,
the magnetic moments are given by~\cite{Fujikawa:1980yx,Pal:1981rm,Shrock:1982sc,Dvornikov:2003js,Dvornikov:2004sj,Giunti:2014ixa}
\begin{equation}
\mu_{kj}^{\text{D}}
\simeq
\frac{3 e G_{\text{F}}}{16\sqrt{2}\pi^{2}}
\left( m_{k} + m_{j} \right)
\left(
\delta_{kj}
-
\frac{1}{2} \sum_{\ell=e,\mu,\tau}
U^{*}_{\ell k} U_{\ell j}
\frac{m_{\ell}^{2}}{m_{W}^{2}}
\right)
,
\label{muD}
\end{equation}
where
$e$ is the elementary charge,
$G_{\text{F}}$ is the Fermi constant,
$m_{k}$ are the neutrino masses for $k=1,2,3$,
$m_{\ell}$ are the charged lepton masses for $\ell=e,\mu,\tau$,
and
$m_{W}$ is the $W$-boson mass.
Note that the magnetic moments are proportional to the neutrino masses,
whose contributions are required in order to flip the chirality in Eq.\ref{dim5lagD},
as illustrated in Fig.\ref{fig:chiralityflip}.
For the diagonal magnetic moments we obtain
\begin{equation}
\mu^{\text{D}}_{kk}
\simeq
\frac{ 3 e G_{\text{F}}  m_{k} }{ 8 \sqrt{2} \pi^{2} }
\simeq
3 \times 10^{-19}
\left( \frac{m_{k}}{\text{eV}} \right) \mu_{\text{B}}
.
\label{muDdiag}
\end{equation}
Therefore,
in the simplest Dirac extension of the SM
the diagonal magnetic moments are strongly suppressed
by the smallness of neutrino masses.
The transition magnetic moments are further suppressed by about
four orders of magnitude:
\begin{equation}
\mu_{kj}^{\text{D}}
\simeq
\frac{3 e G_{\text{F}}}{32\sqrt{2}\pi^{2}}
\left( m_{k} + m_{j} \right)
\sum_{\ell=e,\mu,\tau}
U^{*}_{\ell k} U_{\ell j}
\frac{m_{\ell}^{2}}{m_{W}^{2}}
\simeq
-
3.9 \times 10^{-23}
\left(\frac{m_{k} + m_{j}}{\text{eV}}\right)
\left(
\sum_{\ell=e,\mu,\tau}
U^{*}_{\ell k} U_{\ell j}
\frac{m_{\ell}^2}{m_{\tau}^2}
\right)
\mu_{\text{B}}
,
\label{muDtran}
\end{equation}
for $k \neq j$.
The last term depending on the mixing slightly decreases the values of the transition magnetic moments:
using the values of the mixing angles in Eq.\ref{mixbf}
in the case of Normal Ordering, we obtain
\begin{equation}
|\mu_{12}^{\text{D}}|
\simeq
1.1 \times 10^{-23}
\left(\frac{m_{1} + m_{2}}{\text{eV}}\right)
\mu_{\text{B}}
,
\quad
|\mu_{13}^{\text{D}}|
\simeq
1.3 \times 10^{-23}
\left(\frac{m_{1} + m_{3}}{\text{eV}}\right)
\mu_{\text{B}}
,
\quad
|\mu_{23}^{\text{D}}|
\simeq
1.4 \times 10^{-23}
\left(\frac{m_{2} + m_{3}}{\text{eV}}\right)
\mu_{\text{B}}
.
\label{muDtranbf}
\end{equation}

In the case of Majorana neutrinos the right-handed fields $\nu_{jR}$
in Eq.\ref{dim5lagD}
are replaced by $\nu_{jL}^{c}$:
\begin{equation}
\mathcal{L}_{\text{mag}}^{\text{M}}
=
- \dfrac{1}{4}
\sum_{k,j}
\overline{\nu_{kL}}
\,
\sigma^{\alpha\beta}
\mu_{kj}^{\text{M}}
\nu_{jL}^{c}
\,
F_{\alpha\beta}
+
\text{h.c.}
,
\label{dim5lagM}
\end{equation}
with an additional factor $1/2$ to avoid double counting.
In the minimal extension of the SM with Majorana neutrino masses
it is necessary to have a Majorana mass contribution in order to flip the chirality
($\nu_{jL}^{c}$ is right-handed)
and the magnetic moments are again proportional to the neutrino masses~\cite{Shrock:1982sc}.
Since the Lagrangian~\ref{dim5lagM}
is antisymmetric with respect to the neutrino mass indices,
Majorana neutrinos can have only imaginary transition magnetic moments
(in agreement with Eq.\ref{ffmaj1}).
They are as suppressed as the Dirac transition magnetic moments
in the minimal extension of the SM with Majorana neutrino masses~\cite{Shrock:1982sc}.
However,
in more elaborate models
the Majorana magnetic moments can be enhanced by several orders of magnitude~\cite{Pal:1981rm,Barr:1990um,Babu:1990wv,Pal:1991qr,Boyarkin:2014oza,Lindner:2017uvt}.

In Tables~\ref{tab:MM-SBL} and~\ref{tab:MM-AST}
we present a list of the bounds on neutrino magnetic moments
obtained from observations.
They concern effective neutrino magnetic moments which are explained in the following.
In any case, one can see that the current sensitivity is several orders of magnitude
lower than that required to test the values in Eqs.\ref{muDdiag}--\ref{muDtranbf}.
Nevertheless,
the bounds in Tables~\ref{tab:MM-SBL} and~\ref{tab:MM-AST} are significant and
it is important to develop new experiments with improved sensitivity
in order to test non-minimal BSM models
which predict large neutrino magnetic moments~\cite{Giunti:2014ixa,Lindner:2017uvt}.

\begin{table}[t!]
\begin{minipage}{\textwidth}
\centering
\renewcommand{\arraystretch}{1.2}
\begin{tabular}{llcrcc}
Method & Experiment & Limit $[\mu_{\text{B}}]$ & CL & Year& Ref.\\
\hline
\multirow{6}{*}{Reactor $\bar\nu_e$ E$\nu$ES}
&Krasnoyarsk		&$\mu_{\nu_e} < 2.4 \times 10^{-10}$	&90\%	&1992&\cite{Vidyakin:1992nf}\\
&Rovno			&$\mu_{\nu_e} < 1.9 \times 10^{-10}$	&95\%	&1993&\cite{Derbin:1993wy}\\
&MUNU			&$\mu_{\nu_e} < 9   \times 10^{-11}$	&90\%	&2005&\cite{MUNU:2005xnz}\\
&TEXONO			&$\mu_{\nu_e} < 7.4 \times 10^{-11}$	&90\%	&2006&\cite{TEXONO:2006xds}\\
&GEMMA			&$\mu_{\nu_e} < 2.9 \times 10^{-11}$	&90\%	&2012&\cite{Beda:2012zz}\\
& CONUS
& $\mu_{\nu_e} < 7.5 \times 10^{-11}$
& 90\%
& 2022
& \cite{CONUS:2022qbb}
\\
\hline
\multirow{1}{*}{Accelerator $\nu_e$ E$\nu$ES}
&LAMPF			&$\mu_{\nu_e} < 1.1 \times 10^{-9}$		&90\%	&1992&\cite{Allen:1992qe}\\
\hline
\multirow{3}{*}{Accelerator $\nu_{\mu},\bar\nu_{\mu}$ E$\nu$ES}
&BNL-E734		&$\mu_{\nu_{\mu}} < 8.5 \times 10^{-10}$	&90\%	&1990&\cite{Ahrens:1990fp}\\
&LAMPF			&$\mu_{\nu_{\mu}} < 7.4 \times 10^{-10}$	&90\%	&1992&\cite{Allen:1992qe}\\
&LSND			&$\mu_{\nu_{\mu}} < 6.8 \times 10^{-10}$	&90\%	&2001&\cite{LSND:2001akn}\\
\hline
\multirow{2}{*}{Accelerator $\nu_{\tau},\bar\nu_{\tau}$ E$\nu$ES}
&BEBC~\cite{BEBCWA66:1986err}	&$\mu_{\nu_{\tau}} < 5.4 \times 10^{-7}$	&90\%	&1991&\cite{Cooper-Sarkar:1991vsl}\\
&DONUT			&$\mu_{\nu_{\tau}} < 3.9 \times 10^{-7}$	&90\%	&2001&\cite{DONUT:2001zvi}
\\
\hline
\begin{tabular}{c}
Accelerator $\nu_{e},\nu_{\mu},\bar\nu_{\mu}$
\\[-0.1cm]
CE$\nu$NS+E$\nu$ES
\end{tabular}
&
COHERENT~\cite{COHERENT:2020iec,COHERENT:2021xmm}
&
$
\begin{array}{c} \displaystyle
\mu_{\nu_e} < 4.2 \times 10^{-9}
\\[-0.1cm] \displaystyle
\mu_{\nu_{\mu}} < 1.8 \times 10^{-9}
\end{array}
$
& 90\%
& 2022
& \cite{Coloma:2022avw,AtzoriCorona:2022qrf}
\\
\hline
Reactor $\bar\nu_{e}$ CE$\nu$NS+E$\nu$ES
& Dresden-II~\cite{Colaresi:2022obx}\footnote{Using the Fef quenching factor.}
& $\mu_{\nu_e} < 2.1 \times 10^{-10}$
& 90\%
& 2022
& \cite{Coloma:2022avw,AtzoriCorona:2022qrf}
\\
\hline
\multirow{2}{*}{$e^{+}e^{-}\to\nu\bar\nu\gamma$ (Dirac $\nu$)}
& \hspace{-0.5cm} ASP, MAC, CELLO, MARK-J
& $\mu_{\nu} < 4 \times 10^{-6}$
& 90\%
& 1988
& \cite{Grotch:1988ac}
\\
& TRISTAN
& $\mu_{\nu} < 8.0 \times 10^{-6}$
& 90\%
& 2000
& \cite{Tanimoto:2000am}
\\
\hline
\end{tabular}

\end{minipage}
\caption{ \label{tab:MM-SBL}
Limits for the neutrino magnetic moments obtained in laboratory short-baseline experiments.
$\mu_{\nu}$
indicates a generic neutrino flavor.
}
\end{table}

\begin{table}[t!]
\begin{minipage}{\textwidth}
\centering
\renewcommand{\arraystretch}{1.2}
\begin{tabular}{llcrcc}
Method & Experiment & Limit $[\mu_{\text{B}}]$ & CL & Year& Ref.\\
\hline
\multirow{11}{*}{Solar E$\nu$ES}
& Super-Kamiokande
& $\mu_{\text{S}}^{\text{HE}} < 1.1 \times 10^{-10}$
& 90\%
& 2004
& \cite{Super-Kamiokande:2004wqk}
\\[-0.35cm]
&
\multicolumn{5}{c}{\rule{0.8\linewidth}{0.5pt}}
\\[-0.1cm]
& Borexino
&
\renewcommand{\arraystretch}{0.8}
$\begin{array}{c} \displaystyle
\mu_{\text{S}}^{\text{LE}} < 2.8 \times 10^{-11}
\\ \displaystyle
\mu_{\nu_{e}} < 3.9 \times 10^{-11}
\\ \displaystyle
\mu_{\nu_{\mu}} , \mu_{\nu_{\tau}} < 5.8 \times 10^{-11}
\end{array}$
& 90\%
& 2017
& \cite{Borexino:2017fbd}
\\[-0.3cm]
&
\multicolumn{5}{c}{\rule{0.8\linewidth}{0.5pt}}
\\[-0.15cm]
& XMASS-I
& $ \mu_{\text{S}}^{\text{LE}} < 1.8 \times 10^{-10}$
& 90\%
& 2020
& \cite{XMASS:2020zke}
\\
& XENONnT
& $\mu_{\text{S}}^{\text{LE}} < 6.4 \times 10^{-12}$
& 90\%
& 2022
& \cite{XENON:2022ltv}
\\
& LUX-ZEPLIN
& $\mu_{\text{S}}^{\text{LE}} < 1.36 \times 10^{-11}$
& 90\%
& 2023
& \cite{LZ:2023poo}
\\
& PandaX-4T
& $\mu_{\text{S}}^{\text{LE}} < 2.2 \times 10^{-11}$
& 90\%
& 2024
& \cite{PandaX:2024cic}
\\[-0.35cm]
&
\multicolumn{5}{c}{\rule{0.8\linewidth}{0.5pt}}
\\[-0.1cm]
& LUX-ZEPLIN~\cite{LZ:2022lsv}
&
\renewcommand{\arraystretch}{0.8}
$\begin{array}{c} \displaystyle
\mu_{\text{S}}^{\text{LE}} < 1.1 \times 10^{-11}
\\ \displaystyle
\mu_{\nu_{e}} < 1.5 \times 10^{-11}
\\ \displaystyle
\mu_{\nu_{\mu}} < 2.3 \times 10^{-11}
\\ \displaystyle
\mu_{\nu_{\tau}} < 2.1 \times 10^{-11}
\end{array}$
& 90\%
& 2022
& \cite{AtzoriCorona:2022jeb}
\\[-0.35cm]
&
\multicolumn{5}{c}{\rule{0.8\linewidth}{0.5pt}}
\\[-0.1cm]
& XENONnT~\cite{XENON:2022ltv}
&
\renewcommand{\arraystretch}{0.8}
$\begin{array}{c} \displaystyle
\mu_{\text{S}}^{\text{LE}} < 6.3 \times 10^{-12}
\\ \displaystyle
\mu_{\nu_{e}} < 8.5 \times 10^{-12}
\\ \displaystyle
\mu_{\nu_{\mu}} < 1.4 \times 10^{-11}
\\ \displaystyle
\mu_{\nu_{\tau}} < 1.2 \times 10^{-11}
\end{array}$
& 90\%
& 2022
& \cite{Khan:2022bel}
\\[-0.35cm]
&
\multicolumn{5}{c}{\rule{0.8\linewidth}{0.5pt}}
\\[-0.1cm]
& XENONnT~\cite{XENON:2022ltv}
&
\renewcommand{\arraystretch}{0.8}
$\begin{array}{c} \displaystyle
\mu_{\nu_{e}} < 9.0 \times 10^{-12}
\\ \displaystyle
\mu_{\nu_{\mu}} < 1.5 \times 10^{-11}
\\ \displaystyle
\mu_{\nu_{\tau}} < 1.3 \times 10^{-11}
\end{array}$
& 90\%
& 2022
& \cite{2208.06415}
\\[-0.35cm]
&
\multicolumn{5}{c}{\rule{0.8\linewidth}{0.5pt}}
\\[-0.15cm]
&
\hspace{-0.29cm}
\begin{tabular}{l}
LUX-ZEPLIN~\cite{LZ:2022lsv} $+$
\\[-0.1cm]
PandaX-4T~\cite{PandaX:2022ood} $+$
\\[-0.1cm]
XENONnT~\cite{XENON:2022ltv}
\end{tabular}
&
\renewcommand{\arraystretch}{0.8}
$\begin{array}{c} \displaystyle
\mu_{\text{S}}^{\text{LE}} < 7.5 \times 10^{-12}
\\ \displaystyle
\mu_{\nu_{e}} < 1.0 \times 10^{-11}
\\ \displaystyle
\mu_{\nu_{\mu}} , \mu_{\nu_{\tau}} < 1.6 \times 10^{-11}
\end{array}
$
& 90\%
& 2023
& \cite{Giunti:2023yha}
\\
\hline
\multirow{3}{*}{\hspace{-0.2cm}\begin{tabular}{c}
Core-Collapse
\\[-0.1cm]
Supernovae
\end{tabular}}
&& $\mu_{\nu} \lesssim (2-8) \times 10^{-12}$	&	&1988&\cite{Barbieri:1988nh}
\\[-0.1cm]
&& $\mu_{\nu} \lesssim (1-4) \times 10^{-12}$	&	&1998&\cite{Ayala:1998qz,Ayala:1999xn}
\\[-0.1cm]
&& $\mu_{\nu} \lesssim (1.1-2.7) \times 10^{-12}$&	&2009&\cite{Kuznetsov:2009zm}
\\
\hline
\multirow{7}{*}{\hspace{-0.2cm}\begin{tabular}{c}
Tip of the Red
\\
Giant Branch
\\
(TRGB)
\end{tabular}}
&& $\mu_{\nu} \lesssim 3 \times 10^{-12}$	&	&1989&\cite{Raffelt:1990pj}
\\[-0.1cm]
&& $\mu_{\nu} \lesssim 1 \times 10^{-12}$	&	&1993&\cite{Castellani:1993hs}
\\[-0.1cm]
&& $\mu_{\nu} < 4.5 \times 10^{-12}$		&95\%	&2013&\cite{Viaux:2013hca}
\\[-0.1cm]
&& $\mu_{\nu} \lesssim 2.6 \times 10^{-12}$	&	&2015&\cite{Arceo-Diaz:2015pva}
\\[-0.1cm]
&& $\mu_{\nu} < 1.2 \times 10^{-12}$		&95\%	&2020&\cite{Capozzi:2020cbu}
\\[-0.1cm]
&& $\mu_{\nu} \lesssim (1-5) \times 10^{-12}$	&	&2020&\cite{Mori:2020qqd}
\\[-0.1cm]
&& $\mu_{\nu} \lesssim 6 \times 10^{-12}$	&	&2023&\cite{Franz:2023gic}
\\
\hline
Solar Cooling
&
& $\mu_{\nu} \lesssim 4 \times 10^{-10}$
&
& 1999
& \cite{Raffelt:1999gv}
\\
\hline
\multirow{1}{*}{Cepheid Stars}
&& $\mu_{\nu} \lesssim 2 \times 10^{-10}$	&	&2020&\cite{Mori:2020niw}
\\
\hline
\multirow{2}{*}{White Dwarfs}
&& $\mu_{\nu} \lesssim (7-9) \times 10^{-12}$	&	&2014&\cite{Corsico:2014mpa}
\\[-0.35cm]
&
\multicolumn{5}{c}{\rule{0.8\linewidth}{0.5pt}}
\\[-0.15cm]
&& $\mu_{\nu} < 5 \times 10^{-12}$	&95\%	&2014&\cite{MillerBertolami:2014oki}
\\
\hline
\multirow{3}{*}{\hspace{-0.2cm}\begin{tabular}{c}
Big-Bang
\\[-0.2cm]
Nucleosynthesis
\\[-0.2cm]
(BBN)
\end{tabular}}
&& $\mu_{\nu} \lesssim (1-2) \times 10^{-11}$	&	&1981&\cite{Morgan:1981zy}
\\[-0.1cm]
&& $\mu_{\nu} \lesssim 6.2 \times 10^{-11}$	&	&1997&\cite{Elmfors:1997tt}
\\[-0.1cm]
&& $\mu_{\nu} \lesssim 4 \times 10^{-12}$	&	&2023&\cite{Grohs:2023xwa}
\\
\hline
\multirow{3}{*}{Cosmological $N_{\text{eff}}$}
&& $\mu_{\nu} < 2.7 \times 10^{-12}$	&95\%	&2022&\cite{Li:2022dkc}
\\[-0.1cm]
&& $\mu_{\nu} < 2.6 \times 10^{-12}$	&95\%	&2022&\cite{Carenza:2022ngg}
\\[-0.1cm]
&& $\mu_{\nu} < 5 \times 10^{-12}$	&68\%	&2023&\cite{Grohs:2023xwa}
\\
\hline
\end{tabular}

\end{minipage}
\caption{ \label{tab:MM-AST}
Astrophysical limits for the neutrino magnetic moments.
$\mu_{\text{S}}^{\text{LE}}$ and $\mu_{\text{S}}^{\text{HE}}$
are the low-energy and high-energy effective
magnetic moments in solar neutrino experiments
(see Eqs.\ref{eq:muflavsun}--\ref{eq:PeeHE}).
$\mu_{\nu}$
indicates a generic neutrino flavor.
}
\end{table}

In laboratory experiments the effects of neutrino magnetic moments
are searched through elastic neutrino-electron scattering (\eves)
or with coherent elastic neutrino-nucleus scattering (\cevns).
The latter method has become available only recently,
after the observation of \cevns in the COHERENT experiment~\cite{COHERENT:2017ipa,COHERENT:2020iec,COHERENT:2021xmm}.

Elastic neutrino-electron scattering of a flavor neutrino $\nu_{\ell}$ is the process
\begin{equation}
\nu_{\ell} + e^{-} \to \nu_{\ell'} + e^{-}
.
\label{eq:eves}
\end{equation}
In the SM the neutrino flavor is conserved
($\ell'=\ell$)
and the \eves cross section
of a flavor neutrino $\nu_{\ell}$
with an electron bound in an atom
$\mathcal{A}$
is given by
\begin{equation}
\dfrac{d\sigma_{\nu_{\ell}-\mathcal{A}}^{\eves,\text{SM}}}{d T_{\text{e}}}
(E_{\nu},T_{\text{e}})
=
Z_{\text{eff}}^{\mathcal{A}}(T_{e})
\,
\dfrac{G_{\text{F}}^2 m_{e}}{2\pi}
\left[
\left( g_{V}^{\nu_{\ell}} + g_{A}^{\nu_{\ell}} \right)^2
+
\left( g_{V}^{\nu_{\ell}} - g_{A}^{\nu_{\ell}} \right)^2
\left( 1 - \dfrac{T_{e}}{E_{\nu}} \right)^2
-
\left( (g_{V}^{\nu_{\ell}})^2 - (g_{A}^{\nu_{\ell}})^2 \right)
\dfrac{m_{e} T_{e}}{E_{\nu}^2}
\right]
,
\label{eq:ES-SM}
\end{equation}
where
$G_{\text{F}}$ is the Fermi constant,
$m_{e}$ is the electron mass,
$E_{\nu}$ is the neutrino energy, and
$T_e$ is the observable electron recoil energy.
The neutrino-electron couplings $g_{V,A}^{\nu_{\ell}}$
depend on the neutrino flavor~\cite{AtzoriCorona:2022jeb}:
\begin{align}
g_{V}^{\nu_{e}}
=
\null & \null
2 \sin^{2}\vartheta_{W} + 1/2 + \text{r.c.}
=
0.9521
,
\null & \null
g_{A}^{\nu_{e}}
=
\null & \null
1/2 + \text{r.c.}
=
0.4938
,
\label{eq:gnue}
\\
g_{V}^{\nu_{\mu}}
=
\null & \null
2 \sin^{2}\vartheta_{W} - 1/2 + \text{r.c.}
=
- 0.0397
,
\null & \null
g_{A}^{\nu_{\mu}}
=
\null & \null
- 1/2 + \text{r.c.}
=
- 0.5062
,
\label{eq:gnumu}
\\
g_{V}^{\nu_{\tau}}
=
\null & \null
2 \sin^{2}\vartheta_{W} - 1/2 + \text{r.c.}
=
- 0.0353
,
\null & \null
g_{A}^{\nu_{\tau}}
=
\null & \null
- 1/2 + \text{r.c.}
=
- 0.5062
,
\label{eq:gnutau}
\end{align}
where
$\sin^{2}\vartheta_{W} = 0.23873 \pm 0.00005$~\cite{ParticleDataGroup:2024cfk}
is the weak mixing angle and
``\text{r.c.}'' are radiative corrections.
The coefficient $Z_{\text{eff}}^{\mathcal{A}}(T_{e})$
quantifies the effective number of electrons which can be ionized at $T_e$~\cite{Fayans:2000ns}
(see also Refs.\citenum{Kouzakov:2014lka,Coloma:2022avw,AtzoriCorona:2022qrf,AtzoriCorona:2022jeb})\footnote{
There are also other more complicated and probably more accurate methods
for the description of the atomic ionization effect,
as
the equivalent photon approximation (EPA)
and
the multi-configuration relativistic random phase approximation (MCRRPA)
(see, e.g., Ref.\citenum{Chen:2013lba}).
}.
For antineutrinos the sign of $g_{A}^{\nu_{\ell}}$ must be changed.

Coherent neutrino-nucleus scattering is the process
(see, e.g., the reviews~\citenum{Abdullah:2022zue,Cadeddu:2023tkp})
\begin{equation}
\nu_{\ell} + {}^{A}_{Z}\mathcal{N}
\to
\nu_{\ell'} + {}^{A}_{Z}\mathcal{N}
,
\label{eq:CEvNSproc}
\end{equation}
with a coherent interaction of the neutrino with the nucleus,
such that the nucleus recoils as a whole,
without any change in its internal structure.
This process was predicted theoretically in 1973~\cite{Freedman:1973yd},
but it was observed experimentally for the first time only in 2017
by the COHERENT experiment~\cite{COHERENT:2017ipa},
because the only observable is the recoil of the nucleus with an extremely small kinetic energy which can be measured only with recently-developed
very sensitive detectors~\cite{Xu:2023rzm}.
The \cevns cross-section
of $\nu_{\ell}$ with a nucleus ${}^{A}_{Z}\mathcal{N}$
with $Z$ protons and $N=A-Z$ neutrons
is given by\footnote{This
cross section is calculated for spin-zero nuclei.
The small corrections
for nuclei with low spin are usually neglected~\cite{Barranco:2005yy}.}
\begin{equation}
\dfrac{d\sigma_{\nu_{\ell}\mathcal{N}}^{\cevns}}{d T_{\mathcal{N}}}(E_{\nu},T_{\mathcal{N}})
=
\dfrac{ G_{\text{F}}^2 M_{\mathcal{N}} }{ \pi }
\left(
1 - \dfrac{ M_{\mathcal{N}} T_{\mathcal{N}} }{ 2 E_{\nu}^2 }
- \dfrac{ T_{\mathcal{N}} }{ E_{\nu} }
\right)
\left[ Q_{\nu_{\ell}\mathcal{N}}(|\vet{q}|) \right]^2
,
\label{eq:CEvNScs}
\end{equation}
where
$T_\mathcal{N}$ is the kinetic recoil energy of the nucleus,
$M_{\mathcal{N}}$ is the nuclear mass,
$Q_{\nu_{\ell}\mathcal{N}}(|\vet{q}|)$
is the charge of the nucleus for the interactions
which contribute to \cevns,
and
$|\vet{q}| \simeq \sqrt{ 2 M_{\mathcal{N}} T_{\mathcal{N}} }$
is the absolute value of the three-momentum 
transferred from the neutrino to the nucleus.
In the SM \cevns is a weak neutral current interaction process mediated by the $Z^{0}$ neutral boson, $\ell'=\ell$, and
\begin{equation}
Q_{\nu_{\ell}\mathcal{N}}^{\text{SM}}(|\vet{q}|)
=
g_{V}^{n}
\,
N
\,
F_{N}^{\mathcal{N}}(|\vet{q}|)
+
g_{V}^{p}(\nu_{\ell})
\,
Z
\,
F_{Z}^{\mathcal{N}}(|\vet{q}|)
\label{eq:SMQW}
\end{equation}
is called ``weak charge of the nucleus''.
The functions
$F_{N}^{\mathcal{N}}(|\vet{q}|)$
and
$F_{Z}^{\mathcal{N}}(|\vet{q}|)$
are, respectively, the neutron and proton form factors of the nucleus ${}^{A}_{Z}\mathcal{N}$,
which are the Fourier transforms of the neutron and proton distributions in the nucleus and
characterize the amount of coherency of the interaction.
The coefficients
$g_{V}^{n}$ and $g_{V}^{p}$
are given by~\cite{AtzoriCorona:2023ktl}
\begin{equation}
g_{V}^{n} = - \frac{1}{2} + \text{r.c.} = - 0.5117
\quad
\text{and}
\quad
g_{V}^{p}(\nu_{\ell})
=
\dfrac{1}{2} - 2 \sin^2\!\vartheta_{W} + \text{r.c.}
=
\left\{
\begin{array}{l} \displaystyle
0.0382
\quad \text{for} \quad \nu_{e},
\\ \displaystyle
0.0300
\quad \text{for} \quad \nu_{\mu},
\\ \displaystyle
0.0256
\quad \text{for} \quad \nu_{\tau}.
\end{array}
\right.
\label{eq:CEvNSgV}
\end{equation}
Therefore,
taking also into account that in typical \cevns heavy nuclear targets
$N>Z$,
the neutron contribution is much larger than the proton contribution
and the SM \cevns cross section is approximately proportional to
the square of the neutron number $N$ of the nucleus.

The cross section of the elastic scattering of neutrinos with a target particle
due to an effective neutrino magnetic moment $\mu_{\nu}$
(discussed in the following)
is given by\footnote{
Note that the dependence of this cross section on the electron mass
independently of the target particle
is only apparent and due to the definition
$\mu_{\text{B}} \equiv e / 2 m_{e}$ of the Bohr magneton.
Indeed,
$ \pi \alpha^{2} / m_{e}^{2} \mu_{\text{B}}^2 = \alpha $.
}
\begin{equation}
\frac{d\sigma(\mu_{\nu})}{dT}
=
C
\,
\frac{\pi\alpha^{2}}{m_{e}^{2}}
\left(\frac{1}{T}-\frac{1}{E_{\nu}}\right)
\frac{\mu_{\nu}^2}{\mu_{\text{B}}^2}
,
\label{MMcs}
\end{equation}
where
$\alpha$ is the fine-structure constant,
$T$ is the observable kinetic energy of the recoil target particle,
and $C$ is a coefficient which depends on the target particle:
$C=1$ and $T=T_{e}$ for elastic scattering with a free electron;
$C=Z_{\text{eff}}^{\mathcal{A}}(T_{e})$ and $T=T_{e}$
for elastic scattering with an electron bound in an atom $\mathcal{A}$;
$ C = Z^2 [F_{Z}^{\mathcal{N}}(|\vet{q}|)]^2 $ and $T=T_{\mathcal{N}}$
in \cevns with a nucleus with atomic number $Z$.

\begin{figure}
\hfill
\begin{minipage}{0.4\textwidth}
\includegraphics[width=\textwidth]{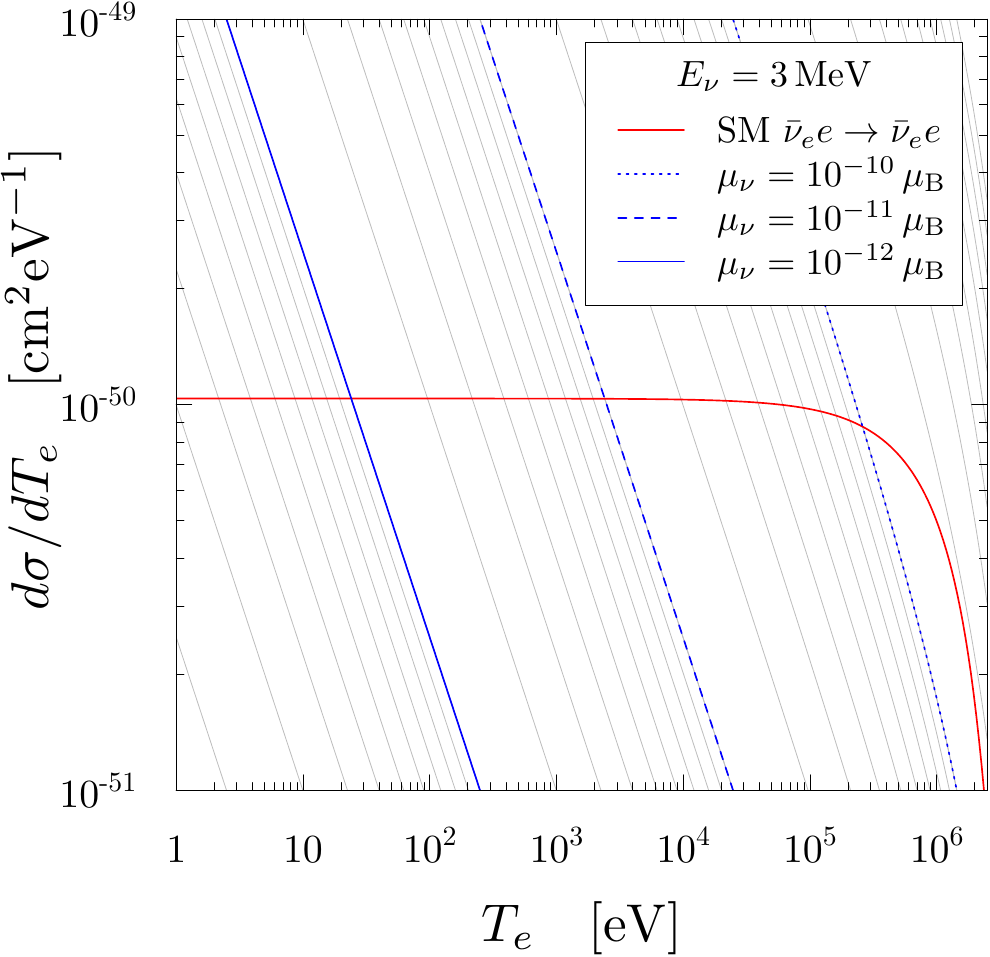}
\end{minipage}
\hfill
\begin{minipage}{0.4\textwidth}
\includegraphics[width=\textwidth]{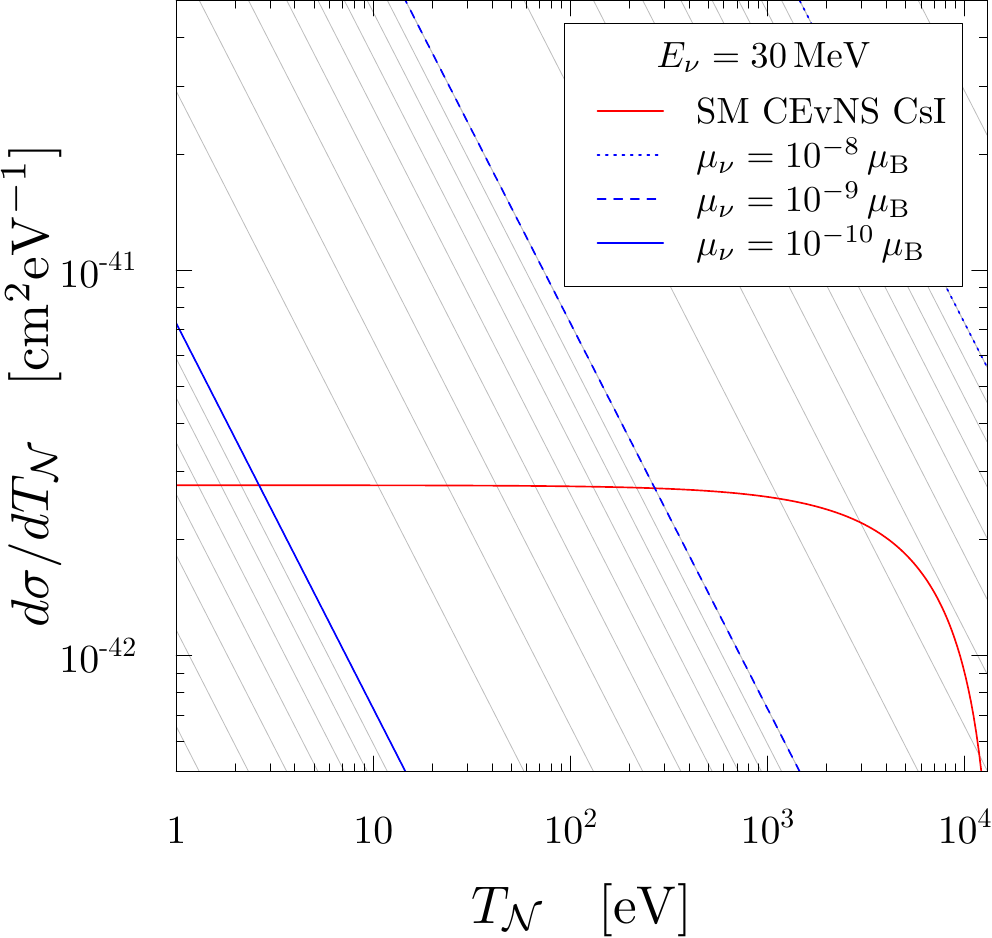}
\end{minipage}
\hfill
\caption{Differential cross sections for
elastic neutrino-electron scattering (\eves, left)
and
coherent neutrino-nucleus scattering (\cevns, right)
as functions of the observable kinetic energy of the recoil target particle.
For \eves the value of the neutrino energy $E_{\nu}$
is typical of reactor neutrino experiments,
whereas for \cevns it is typical of neutrinos produced by pion decay at rest,
as in the COHERENT experiment~\cite{COHERENT:2017ipa}.
For \cevns the cross sections are the average cross sections for
$^{133}\text{Cs}$ and $^{127}\text{I}$,
which constitute the COHERENT CsI detector~\cite{COHERENT:2017ipa}.
}
\label{fig:magfig}
\end{figure}

Since the magnetic moment interaction flips
the helicity of ultrarelativistic neutrinos,
it is not coherent with the SM weak interaction
and the total cross section of a scattering process is the sum of the two cross sections.
As shown in Fig.\ref{fig:magfig},
the magnetic moment cross section can be larger than the weak cross section
for low values of $T$, because of its approximate $1/T$ dependence.
When this happens it distorts the energy spectrum of the recoil target particle,
giving information on the value of the magnetic moment.
From Fig.\ref{fig:magfig}
one can see that both \eves and \cevns
are sensitive to the neutrino magnetic moment,
but \eves of reactor neutrinos can probe smaller values of $\mu_{\nu}$
than \cevns in the COHERENT experiment.
Indeed,
one can see from Table~\ref{tab:MM-SBL}
that \eves experiments with reactor $\bar{\nu}_{e}$'s
have obtained bounds on the effective
$\mu_{\nu_{e}}$ which are better than the COHERENT bound
(for which a small \eves contribution, which is indistinguishable from the main \cevns signal in the COHERENT detector, has been taken into account).
The most stringent bound on $\mu_{\nu_{e}}$ has been obtained
by the GEMMA~\cite{Beda:2012zz} experiment from \eves of reactor $\bar{\nu}_{e}$'s
at the Kalinin Nuclear Power Plant in Russia,
with a 1.5 kg high-purity germanium (HPGe) detector with an energy threshold of 2.8~keV.
On the other hand,
since the neutrino beam in the COHERENT experiment
includes $\nu_{\mu}$'s and $\bar{\nu}_{\mu}$'s,
the COHERENT data constrain also
$\mu_{\nu_{\mu}}$.
From Table~\ref{tab:MM-SBL},
one can see that the \cevns+\eves COHERENT bound on
$\mu_{\nu_{\mu}}$
is not far from the more stringent bounds obtained in accelerator \eves experiments.

\red{In Table~\ref{tab:MM-SBL}
we report also generic limits on the neutrino magnetic moments
obtained from searches of $e^+ e^- \to \nu \bar\nu \gamma$ events
in high-energy experiments
neglecting a possible $q^2$ dependence of the magnetic form factor.}

The magnetic moment in Eq.\ref{MMcs}
is an effective magnetic moment~\cite{Grimus:1997aa,Beacom:1999wx}
which depends on the experiment under consideration and is given, in general, by
\begin{equation}
\mu_{\nu}^2
=
\sum_{j}
\left|
\sum_{k}
\mu_{jk}
A_{k}(E_{\nu},L)
\right|^2
,
\label{mueff}
\end{equation}
where $A_{k}(E_{\nu},L)$
is the amplitude of the massive neutrino $\nu_{k}$
with energy $E_{\nu}$
after propagation over a source-detector distance $L$.
The contributions of the
detected initial massive neutrinos in the interaction process,
labeled with the index $k$,
are summed coherently,
whereas the contributions of the undetected
final massive neutrinos in the interaction process,
labeled with the index $j$,
are summed incoherently.
 
For neutrino propagation in vacuum
$A_{k}(E_{\nu},L) = A_{k}^{0}(E_{\nu}) e^{-i m_{k}^2 L / 2 E_{\nu}}$,
where
$A_{k}^{0}(E_{\nu})$
is the amplitude of $\nu_{k}$ production at the source
and
$e^{-i m_{k}^2 L / 2 E_{\nu}}$
is the phase of $\nu_{k}$.
If the source emits a $\nu_{\ell}$ or $\bar\nu_{\ell}$ flux
in matter with a normal density
$A_{k}^{0} \simeq U_{\ell k}^{*}$
or
$A_{k}^{0} \simeq U_{\ell k}$,
respectively.

In short-baseline experiments with neutrinos from reactors and accelerators,
the neutrino source-detector distance is shorter than the oscillation length
($L \ll 4 \pi E_{\nu} / \Delta{m}^{2}_{kj}$)
and
the detected neutrino flavor is the same as the flavor of the neutrino emitted by the source
(e.g. $\bar\nu_{e}$ for reactor neutrinos).
In this case,
$A_{k} \simeq U_{\ell k}^{*}$
if the source emits a $\nu_{\ell}$ flux,
and
$A_{k} \simeq U_{\ell k}$
if the source emits a $\bar\nu_{\ell}$ flux.
Therefore,
taking into account the property~\ref{anti1},
the squared effective magnetic moments of $\nu_{\ell}$ and $\bar\nu_{\ell}$
are equally given by
\begin{equation}
\mu_{\nu_{\ell}}^2
\simeq
\mu_{\bar\nu_{\ell}}^2
\simeq
\sum_{j}
\left|
\sum_{k}
\mu_{jk}
U_{\ell k}^{*}
\right|^2
=
\sum_{\ell'}
\left| \mu_{\ell'\ell} \right|^2
=
\mu_{\ell\ell}^2
+
\sum_{\ell'\neq\ell}
\left| \mu_{\ell\ell'} \right|^2
,
\label{eq:muflav}
\end{equation}
where
$\mu_{\ell'\ell}=\mu_{\ell\ell'}^{*}$
are the flavor magnetic moments given by
the transformation~\ref{eq:flavff}:
\begin{equation}
\mu_{\ell'\ell}
=
\sum_{j,k}
U_{\ell'j}
\mu_{jk}
U_{\ell k}^{*}
.
\label{eq:flavmu}
\end{equation}
Therefore,
the squared effective magnetic moment
that is measured in a $\nu_{\ell}$ short-baseline experiment
is given by the sum of the squared diagonal magnetic moment $\mu_{\ell\ell}$
of $\nu_{\ell}$
and the squared absolute values of the transition magnetic moments
$\mu_{\ell\ell'}$ for $\ell'\neq\ell$.
The same result could have been obtained in the flavor basis summing incoherently
over the possible flavors $\ell'$ of the undetected final neutrino $\nu_{\ell'}$
(see Eqs.\ref{eq:eves} and~\ref{eq:CEvNSproc}):
\begin{equation}
\mu_{\nu}^2
=
\sum_{\ell'}
\left|
\sum_{\ell''}
\mu_{\ell'\ell''}
A_{\ell''}(E_{\nu},L)
\right|^2
.
\label{eq:flavmueff}
\end{equation}
In a short-baseline experiment with a $\nu_{\ell}$ or $\bar\nu_{\ell}$ source,
$A_{\ell''}(E_{\nu},L)=\delta_{\ell\ell''}$
and Eq.\ref{eq:flavmueff} gives Eq.\ref{eq:muflav}.

The effective flavor magnetic moments
$\mu_{\nu_{\ell}}$
in Table~\ref{tab:MM-SBL} which are bounded by short-baseline reactor and accelerator neutrino experiments are given by the square root of Eq.\ref{eq:muflav}.
Therefore they constrain not only the diagonal magnetic moment $\mu_{\ell\ell}$
of $\nu_{\ell}$, but also the
transition magnetic moments
$\mu_{\ell\ell'}$ for $\ell'\neq\ell$.

If the neutrino source and detector are far away,
the detected neutrino flux may contain flavor neutrinos which are different from the initial one and
it is convenient to consider the expression~\ref{eq:flavmueff}
of $\mu_{\nu}^2$ in the flavor basis.
In the case of solar neutrinos,
$L$ is very large and
the interference terms in Eq.\ref{eq:flavmueff}
are washed out by the neutrino energy spectrum,
leading to the effective magnetic moment
\begin{equation}
\mu_{\text{S}}^2
\simeq
\sum_{\ell}
P^{\text{S}}_{e\ell}(E_{\nu})
\sum_{\ell'}
\left| \mu_{\ell'\ell} \right|^2
,
\label{eq:solmueff}
\end{equation}
where
$P^{\text{S}}_{e\ell}(E_{\nu})$
is the probability of solar
$ \nu_{e} \to \nu_{\ell} $
transitions.
Taking into account that
$ P^{\text{S}}_{e\mu} = \cos^2\vartheta_{23} \left( 1 - P^{\text{S}}_{ee} \right) $
and
$ P^{\text{S}}_{e\tau} = \sin^2\vartheta_{23} \left( 1 - P^{\text{S}}_{ee} \right) $,
Eq.\ref{eq:solmueff} can be written as
\begin{equation}
\mu_{\text{S}}^2
=
P^{\text{S}}_{ee} \, \mu_{\nu_{e}}^2
+
\left( 1 - P^{\text{S}}_{ee} \right)
\left(
\cos^2\vartheta_{23} \, \mu_{\nu_{\mu}}^2
+
\sin^2\vartheta_{23} \, \mu_{\nu_{\tau}}^2
\right)
.
\label{eq:muflavsun}
\end{equation}
with
$\mu_{\nu_{e}}$,
$\mu_{\nu_{\mu}}$, and
$\mu_{\nu_{\tau}}$
given by Eq.\ref{eq:muflav}.
For low-energy and high-energy solar neutrinos
\begin{alignat}{2}
&
P^{\text{S}}_{ee}
\simeq
\left( 1 - \frac{1}{2} \, \sin^2 2\vartheta_{12} \right)
\cos^4\vartheta_{13}
+
\sin^4\vartheta_{13}
=
P^{\text{S,LE}}_{ee}
&&
\quad
\text{for}
\quad
E_{\nu} \lesssim 1 \, \text{MeV}
,
\label{eq:PeeLE}
\\
&
P^{\text{S}}_{ee}
\simeq
\sin^2 \vartheta_{12}
\cos^4\vartheta_{13}
+
\sin^4\vartheta_{13}
=
P^{\text{S,HE}}_{ee}
&&
\quad
\text{for}
\quad
E_{\nu} \gtrsim 5 \, \text{MeV}
.
\label{eq:PeeHE}
\end{alignat}
Using the values of the mixing angles in Eq.\ref{mixbf},
we obtain
$ P^{\text{S,LE}}_{ee} = 0.542 \pm 0.011 $
and
$ P^{\text{S,HE}}_{ee} = 0.305 \pm 0.015 $
independently of the ordering of the neutrino masses.
Therefore,
the effective solar neutrino magnetic moment obtained
from the analyses of the data of low-energy
($\mu_{\text{S}}^{\text{LE}}$)
and high-energy
($\mu_{\text{S}}^{\text{HE}}$)
solar neutrino experiments
are different and cannot be compared directly.
The results on the measurement of the effective
solar neutrino magnetic moment can be compared directly
only among solar neutrino experiments
which detect neutrinos with the same energy range,
as those for $\mu_{\text{S}}^{\text{LE}}$
in Table~\ref{tab:MM-AST}.
The results on
$\mu_{\nu_{e}}$,
$\mu_{\nu_{\mu}}$, and
$\mu_{\nu_{\tau}}$
can be compared with the caveat that they may have been obtained with
slightly different values of $P^{\text{S}}_{ee}$ and $\vartheta_{23}$
and different treatments of the systematic uncertainties.
In any case,
one can see from Table~\ref{tab:MM-AST} that the best solar bounds on
$\mu_{\nu_{e}}$
and all the solar bounds on
$\mu_{\nu_{\mu}}$ and $\mu_{\nu_{\tau}}$
are stronger than those of reactors and accelerators in Table~\ref{tab:MM-SBL}.

In Table~\ref{tab:MM-AST} we listed also astrophysical bounds
which have larger systematic uncertainties and are less accurate than the previous bounds.
Nevertheless, it is useful to have a view of them, because it is an active field of research and they are quite stringent.
Since these limits
have been obtained without taking into account neutrino mixing,
they apply approximately to all neutrino flavors.
The Core-Collapse Supernovae bounds stem from the limit on a possible energy loss due to scattering processes which induce
magnetic moment transitions from left-handed neutrinos to
sterile right-handed neutrinos which escape from the environment.
The TRGB, Solar Cooling, Cepheid Stars, and White Dwarfs bounds ensue from the energy loss due to plasmon decay
into a neutrino-antineutrino pair~\cite{Raffelt:1996wa}
discussed in Section~\ref{sec:Plasmon}.
The BBN (Big Bang Nucleosynthesis)
and $N_{\text{eff}}$ bounds follow from the constraints on the production of right-handed neutrinos by the scattering of left-handed neutrinos with a magnetic moment with the charged particles in the primordial plasma.
In the calculations of Ref.\cite{Carenza:2022ngg}
the bound on the neutrino magnetic moment depends on the assumed value
$T_{\text{in}}$
of the temperature of the Universe at which the production of right-handed neutrino starts\footnote{$T_{\text{in}}$
may be the energy scale of the new physics which generates the neutrino magnetic moment~\cite{Carenza:2022ngg}.
}.
The bound
$\mu_{\nu} < 2.6 \times 10^{-12}$ at 95\% CL
reported in Table~\ref{tab:MM-AST} was obtained assuming
$T_{\text{in}} = 100 \, \text{GeV}$
and is in approximate agreement with those of Refs.\cite{Li:2022dkc,Grohs:2023xwa},
also reported in Table~\ref{tab:MM-AST}.
Assuming a smaller initial temperature
the limit becomes less stringent,
since less right-handed neutrino are produced
(e.g., for $T_{\text{in}} = 100 \, \text{MeV}$,
the bound becomes
$\mu_{\nu} < 1.6 \times 10^{-11}$ at 95\% CL).
Future possibilities to measure very small neutrino 
magnetic moments with astrophysical neutrinos
have been discussed in Refs.\cite{Kopp:2022cug,Brdar:2023cub}.

\section{Neutrino charge radius}
\label{sec:CR}

The neutrino charge radii describe virtual charge distributions
which can exist even if the neutrinos are neutral.
They are given by the first-order $q^{2}$ dependence of the charge form factors
in Eq.\ref{eq:FQ}.
Therefore, they are phenomenological quantities which can be measured
in neutrino scattering with charged particles,
as \eves and \cevns.

Theoretically,
the neutrino charge radii are generated by radiative corrections
and exist even in the SM,
where neutrinos are neutral and massless~\cite{Bernabeu:2000hf,Bernabeu:2002nw,Bernabeu:2002pd,Erler:2013xha}.
Since in the SM
the generation lepton numbers are conserved,
the SM charge radii are defined in the flavor basis.
Taking into account one-loop radiative corrections, they are given
by~\cite{Bernabeu:2002pd}\footnote{
The difference by a factor of 2 with respect to
the definition of the charge radius
in Ref.\cite{Bernabeu:2002pd}
is explained in Ref.\cite{Cadeddu:2018dux}.
}
\begin{equation}
\langle r_{\nu_{\ell}}^2 \rangle^{\text{SM}}
=
-
\frac{G_{\text{F}}}{2\sqrt{2}\pi^{2}}
\left[
3-2\ln\left(\frac{m_{\ell}^{2}}{m^{2}_{W}}\right)
\right]
=
\left\{
\begin{array}{rl} \displaystyle
- 0.83 \times 10^{-32} \, \text{cm}^{2}
& \displaystyle
(\nu_{e})
\\ \displaystyle
- 0.48 \times 10^{-32} \, \text{cm}^{2}
& \displaystyle
(\nu_{\mu})
\\ \displaystyle
- 0.30 \times 10^{-32} \, \text{cm}^{2}
& \displaystyle
(\nu_{\tau})
\end{array}
\right.
\label{eq:SMchr}
\end{equation}
where $m_{W}$ is the $W$ boson mass
and $m_{\ell}$ are the charged lepton masses for $\ell=e,\mu,\tau$.

Finding deviations of the neutrino charge radii from the SM values
would be a discovery of new BSM physics.
Therefore, we consider the general case of diagonal
and off-diagonal (transition) charge radii in the flavor basis,
which are given, from Eq.\ref{eq:flavff}, by
\begin{equation}
\langle r_{\nu_{\ell'\ell}}^2 \rangle
=
\sum_{k,j}
U_{\ell' k}
\langle r_{\nu_{kj}}^2 \rangle
U_{\ell j}^{*}
.
\label{eq:chrflavor}
\end{equation}
As already noted after Eq.\ref{eq:flavff},
the unitary transformation from the mass to the flavor basis
preserves the hermiticity of the matrix of charge radii:
$\langle r_{\nu_{\ell'\ell}}^2 \rangle=\langle r_{\nu_{\ell\ell'}}^2 \rangle^{*}$.
Hence, the diagonal charge radii of flavor neutrinos
$\langle r_{\nu_{e}}^2 \rangle\equiv\langle r_{\nu_{ee}}^2 \rangle$,
$\langle r_{\nu_{\mu}}^2 \rangle\equiv\langle r_{\nu_{\mu\mu}}^2 \rangle$, and
$\langle r_{\nu_{\tau}}^2 \rangle\equiv\langle r_{\nu_{\tau\tau}}^2 \rangle$
are real.

The traditional way to obtain limits on
$\langle r_{\nu_{e}}^2 \rangle$
is the observation of elastic neutrino-electron scattering (\eves)
using the intense fluxes of reactor $\bar\nu_e$'s.
The limits on
$\langle r_{\nu_{\mu}}^2 \rangle$
have been obtained with \eves
of accelerator $\nu_\mu$'s produced mainly by pion decay.
These methods have been recently complemented by
the measurement of coherent elastic neutrino-nucleus scattering (\cevns)
of low-energy electron and muon neutrinos
produced by pion and muon decays at rest
or reactor $\bar\nu_e$'s.

The total \eves cross section
of a flavor neutrino $\nu_{\ell}$
with an electron bound in an atom
$\mathcal{A}$
is given by~\cite{Kouzakov:2017hbc}
\begin{equation}
\dfrac{d\sigma_{\nu_{\ell}-\mathcal{A}}^{\eves,\text{SM+CR}}}{d T_{\text{e}}}
=
\left(
\dfrac{d\sigma_{\nu_{\ell}-\mathcal{A}}^{\eves,\text{SM+CR}}}{d T_{\text{e}}}
\right)_{\langle{r}_{\nu_{\ell}}^2\rangle}
+
\sum_{\ell'\neq\ell}
\left(
\dfrac{d\sigma_{\nu_{\ell}-\mathcal{A}}^{\eves,\text{CR}}}{d T_{\text{e}}}
\right)_{\langle{r}_{\nu_{\ell\ell'}}^2\rangle}
.
\label{eq:ES-CR}
\end{equation}
The first term takes into account coherently the effects of the
SM weak interaction and the diagonal flavor-conserving neutrino charge radii.
It is obtained from Eq.\ref{eq:ES-SM}
with the substitution
\begin{equation}
g_{V}^{\nu_{\ell}}
\to
g_{V}^{\nu_{\ell}}
+
\dfrac{ \sqrt{2} \pi \alpha }{ 3 G_{\text{F}} }
\left(
\langle{r}_{\nu_{\ell}}^2\rangle
-
\langle{r}_{\nu_{\ell}}^2\rangle^{\text{SM}}
\right)
=
\left\{
\begin{array}{l} \displaystyle
\tilde{g}_{V}^{\nu_{e}}
+
\dfrac{ \sqrt{2} \pi \alpha }{ 3 G_{\text{F}} }
\,
\langle{r}_{\nu_{e}}^2\rangle
\quad
\text{for}
\quad
\ell=e
,
\\ \displaystyle
\tilde{g}_{V}^{\nu_{\mu}}
+
\dfrac{ \sqrt{2} \pi \alpha }{ 3 G_{\text{F}} }
\,
\langle{r}_{\nu_{\ell}}^2\rangle
\quad
\text{for}
\quad
\ell=\mu,\tau
,
\end{array}
\right.
\label{eq:gV-CR}
\end{equation}
where
$ \tilde{g}_{V}^{\nu_{\mu}} = - 0.028 $
and
$ \tilde{g}_{V}^{\nu_{e}}
= \tilde{g}_{V}^{\nu_{\mu}} + 1
= 0.972 $\footnote{Since the other radiative corrections are flavor-independent,
The value of $\tilde{g}_{V}^{\nu_{e}}$
is obtained by adding 1 to $\tilde{g}_{V}^{\nu_{\mu}}$,
where the 1 is due to the charged-current
contribution in $\nu_{e}$ \eves~\cite{ParticleDataGroup:2024cfk}.}.
Since we are interested in the measurement of the neutrino charge radii
including the SM contributions in Eq.\ref{eq:SMchr},
we subtracted from the SM values of
$g_{V}^{\nu_{\ell}}$ given by Eqs.\ref{eq:gnue}--\ref{eq:gnutau}
the contributions of the SM charge radii
which are included in the SM radiative corrections.
Note that $\nu_{\ell}$ \eves experiments
are sensitive to the sign\footnote{This behavior can be compared with the impossibility to measure the sign of the magnetic moment,
which appears as squared in the cross section~\ref{MMcs}.}
of the diagonal charge radius
$\langle{r}_{\nu_{\ell}}^2\rangle = \langle{r}_{\nu_{\ell\ell}}^2\rangle$.
The second term in Eq.\ref{eq:ES-CR}
takes into account the incoherent contribution of the transition charge radii:
\begin{equation}
\left(
\dfrac{d\sigma_{\nu_{\ell}-\mathcal{A}}^{\eves,\text{CR}}}{d T_{\text{e}}}
\right)_{\langle{r}_{\nu_{\ell\ell'}}^2\rangle}
=
Z_{\text{eff}}^{\mathcal{A}}(T_{e})
\,
\dfrac{\pi \alpha^2 m_{e}}{9}
\left[
1
+
\left( 1 - \dfrac{T_{e}}{E_{\nu}} \right)^2
-
\dfrac{m_{e} T_{e}}{E_{\nu}^2}
\right]
|\langle{r}_{\nu_{\ell\ell'}}^2\rangle|^2
\quad \text{for} \quad \ell'\neq\ell
.
\label{eq:ES-CR-tr}
\end{equation}
Therefore,
\eves experiments with a pure $\nu_{\ell}$ beam
are sensitive only to the sum of the squared absolute values of
the transition charge radii
$\langle{r}_{\nu_{\ell\ell'}}^2\rangle$
for $\ell'\neq\ell$~\cite{Kouzakov:2017hbc}.

\begin{table}[t!]
\hspace{-2cm}
\begin{minipage}{\textwidth}
\centering
\renewcommand{\arraystretch}{1.2}
\begin{tabular}{llcrcc}
Method & Experiment & Limit $[10^{-32}\,\text{cm}^2]$ & CL & Year & Ref.\\
\hline
\multirow{2}{*}{Reactor $\bar\nu_e$ E$\nu$ES}
&Krasnoyarsk	&$|\langle r_{\nu_{e}}^2 \rangle|<7.3$		&90\%	&1992&
\cite{Vidyakin:1992nf}\\
&TEXONO		&$\langle r_{\nu_{e}}^2 \rangle\in(-4.2,6.6)$	&90\%	&2009&
\cite{TEXONO:2009knm}~\footnote{\label{f2}Corrected by a factor of two due to a different convention.}\\
\hline
\multirow{2}{*}{Accelerator $\nu_e$ E$\nu$ES}
&LAMPF		&$\langle r_{\nu_{e}}^2 \rangle\in(-7.12,10.88)$	&90\%	&1992&
\cite{Allen:1992qe}~\footref{f2}\\
&LSND		&$\langle r_{\nu_{e}}^2 \rangle\in(-5.94,8.28)$	&90\%	&2001&
\cite{LSND:2001akn}~\footref{f2}
\\
\hline
\multirow{1}{*}{$\nu_{\mu}$-nucleon DIS}
& NuTeV~\cite{NuTeV:2001whx}
& $|\langle r_{\nu_{\mu}}^2 \rangle| < 0.71$
& 90\%
& 2002
& \cite{Hirsch:2002uv}
\\
\hline
\multirow{3}{*}{Accelerator $\nu_{\mu}$ E$\nu$ES}
& BNL-E734~\cite{Ahrens:1990fp}
& $\langle r_{\nu_{\mu}}^2 \rangle \in (-1.1,5.7)$
& 90\%
& 2002
& \cite{Hirsch:2002uv}~\footnote{\label{Hirsch}With inverted sign due to the opposite sign convention.}
\\
& CHARM-II~\cite{CHARM-II:1994aeb}
& $\langle r_{\nu_{\mu}}^2 \rangle \in (-2.2,0.52)$
& 90\%
& 2002
& \cite{Hirsch:2002uv}~\footref{Hirsch}
\\
& CCFR~\cite{CCFR:1997zzq}
& $\langle r_{\nu_{\mu}}^2 \rangle \in (-0.68,0.53)$
& 90\%
& 2002
& \cite{Hirsch:2002uv}~\footref{Hirsch}
\\
\hline
\setlength{\tabcolsep}{0pt}
\hspace{-0.2cm}
\begin{tabular}{c}
Accelerator $\nu_{e},\nu_{\mu},\bar\nu_{\mu}$
\\[-0.1cm]
$+$
Reactor $\bar\nu_{e}$ CE$\nu$NS
\end{tabular}
&
\hspace{-0.29cm}
\begin{tabular}{c}
COHERENT~\cite{COHERENT:2020iec,COHERENT:2021xmm}
\\[-0.1cm]
$+$
Dresden-II~\cite{Colaresi:2022obx}\footnote{Using the Fef quenching factor.}
\end{tabular}
&
$\begin{array}{c} \displaystyle
\langle r_{\nu_{e}}^2 \rangle \in (-7.1,5)
\\ \displaystyle
\langle r_{\nu_{\mu}}^2 \rangle \in (-5.9,4.3)
\end{array}$
& 90\%
& 2022
& \cite{AtzoriCorona:2022qrf}
\\
\hline
\multirow{5}{*}{Solar E$\nu$ES}
& XENONnT~\cite{XENON:2022ltv}
&
$\begin{array}{c} \displaystyle
\langle r_{\nu_{e}}^2 \rangle \in (-85, 2.0)
\\ \displaystyle
\langle r_{\nu_{\mu}}^2 \rangle \in (-45, 52)
\\ \displaystyle
\langle r_{\nu_{\tau}}^2 \rangle \in (-40, 45)
\end{array}$
& 90\%
& 2022
& \cite{Khan:2022bel}
\\[-0.35cm]
&
\multicolumn{5}{c}{\rule{0.8\linewidth}{0.5pt}}
\\[-0.15cm]
& XENONnT~\cite{XENON:2022ltv}
&
$\begin{array}{c} \displaystyle
\langle r_{\nu_{e}}^2 \rangle \in (-93.4, 9.5)
\\ \displaystyle
\langle r_{\nu_{\mu}}^2 \rangle \in (-50.2, 54)
\\ \displaystyle
\langle r_{\nu_{\tau}}^2 \rangle \in (-43, 46.8)
\end{array}$
& 90\%
& 2022
& \cite{2208.06415}
\\[-0.35cm]
&
\multicolumn{5}{c}{\rule{0.8\linewidth}{0.5pt}}
\\[-0.15cm]
&
\hspace{-0.29cm}
\begin{tabular}{c}
LUX-ZEPLIN~\cite{LZ:2022lsv} $+$
\\[-0.1cm]
PandaX-4T~\cite{PandaX:2022ood} $+$
\\[-0.1cm]
XENONnT~\cite{XENON:2022ltv}
\end{tabular}
&
$\begin{array}{c} \displaystyle
\langle r_{\nu_{e}}^2 \rangle \in (-99.5, 12.8)
\\ \displaystyle
\langle r_{\nu_{\mu}}^2 \rangle
,
\langle r_{\nu_{\tau}}^2 \rangle \in (-82.2, 88.7)
\end{array}$
& 90\%
& 2023
& \cite{Giunti:2023yha}
\\
\hline
\multirow{3}{*}{$e^{+}e^{-}\to\nu\bar\nu\gamma$ (Dirac $\nu$)}
& \hspace{-0.5cm} ASP, MAC, CELLO, MARK-J
& $|\langle r_{\nu}^2 \rangle| < 20$
& 90\%
& 1988
& \cite{Grotch:1988ac}
\\
& TRISTAN
& $\langle r_{\nu}^2 \rangle \in (-18,21)$
& 90\%
& 2002
& \cite{Hirsch:2002uv}~\footref{Hirsch}
\\
& LEP-2
& $\langle r_{\nu}^2 \rangle \in (-6.2,5.6)$
& 90\%
& 2002
& \cite{Hirsch:2002uv}~\footref{Hirsch}
\\
\hline
\multirow{2}{*}{$e^{+}e^{-}\to\nu\bar\nu\gamma$ (Majorana $\nu$)}
& TRISTAN
& $\langle r_{\nu}^2 \rangle \in (-31,37)$
& 90\%
& 2002
& \cite{Hirsch:2002uv}~\footref{Hirsch}
\\
& LEP-2
& $\langle r_{\nu}^2 \rangle \in (-9.9,8.2)$
& 90\%
& 2002
& \cite{Hirsch:2002uv}~\footref{Hirsch}
\\
\hline
\multirow{1}{*}{BBN (Dirac $\nu$)}
&
& $|\langle r_{\nu}^2 \rangle| \lesssim 0.7$
&
& 1987
& \cite{Grifols:1986ed}
\\
\hline
\multirow{1}{*}{SN1987A (Dirac $\nu$)}
&
& $|\langle r_{\nu}^2 \rangle| \lesssim 0.2$
&
& 1989
& \cite{Grifols:1989vi}
\\
\hline
\multirow{1}{*}{TRGB}
&
& $|\langle r_{\nu}^2 \rangle| \lesssim 40-50$
&
& 1994
& \cite{Altherr:1993hb}
\\
\hline
\end{tabular}

\end{minipage}
\caption{ \label{tab:CR}
Limits for the charge radii of flavor neutrinos obtained assuming
the absence of transition charge radii.
These limits can be compared with the Standard Model predictions
in Eq.\ref{eq:SMchr}.
$\langle r_{\nu}^2 \rangle$
indicates a generic neutrino flavor.
}
\end{table}

The \cevns cross section of a flavor neutrino $\nu_{\ell}$
with a nucleus ${}^{A}_{Z}\mathcal{N}$
is given by Eq.\ref{eq:CEvNScs} with
\begin{equation}
\left| Q_{\nu_{\ell}\mathcal{N}}^{\text{SM+CR}}(|\vet{q}|) \right|^2
=
\left[
g_{V}^{n}
\,
N
\,
F_{N}^{\mathcal{N}}(|\vet{q}|)
+
\left(
\tilde{g}_{V}^{p}
-
\widetilde{Q}_{\ell\ell}
\right)
Z
\,
F_{Z}^{\mathcal{N}}(|\vet{q}|)
\right]^2
+
\left[
Z
\,
F_{Z}^{\mathcal{N}}(|\vet{q}|)
\right]^2
\sum_{\ell'\neq\ell}
|\widetilde{Q}_{\ell\ell'}|^2
,
\label{eq:CRQW}
\end{equation}
where $\tilde{g}_{V}^{p}=0.0186$
is the neutrino-proton coupling without the contribution of the SM neutrino
charge radius,
which is included in the radiative corrections in Eq.\ref{eq:CEvNSgV}.
The effects of the charge radii
$\langle{r}_{\nu_{\ell\ell'}}^2\rangle$
are given by
\begin{equation}
\widetilde{Q}_{\ell\ell'}
=
\dfrac{ \sqrt{2} \pi \alpha }{ 3 G_{\text{F}} }
\, \langle{r}_{\nu_{\ell\ell'}}^2\rangle
.
\label{eq:Qchr}
\end{equation}
As in the case of \eves,
the contribution of the diagonal charge radius
$\langle{r}_{\nu_{\ell}}^2\rangle = \langle{r}_{\nu_{\ell\ell}}^2\rangle$
is added coherently to the SM weak interaction
and $\nu_{\ell}$ \cevns experiments are sensitive to its sign.
The contributions of the transition charge radii
$\langle{r}_{\nu_{\ell\ell'}}^2\rangle$
(for $\ell'\neq\ell$)
are added incoherently
and $\nu_{\ell}$ \cevns experiments
are sensitive only to the sum of their squared absolute values,
as $\nu_{\ell}$ \eves experiments.

Experiments which have beams of different neutrino flavors
with different characteristics (as the energy spectrum) can
distinguish the contributions
of the different diagonal charge radii
and those of the different transition radii.
This happens in the COHERENT experiment which has
$\nu_{e}$,
$\nu_{\mu}$, and
$\bar\nu_{\mu}$,
beams with different energy and time distributions.
Since
$|\langle{r}_{\nu_{e\mu}}^2\rangle| = |\langle{r}_{\nu_{\mu e}}^2\rangle|$
contributes to all events,
$\langle{r}_{\nu_{e\tau}}^2\rangle$
contributes only to $\nu_{e}$-induced events, and
$\langle{r}_{\nu_{\mu\tau}}^2\rangle$
contributes only to $\nu_{\mu}$ and $\bar\nu_{\mu}$-induced events,
they can be distinguished.
Solar neutrino experiments can distinguish all neutrino charge radii,
because they detect different fluxes of $\nu_{e}$
and of
$\nu_{\mu}$ and $\nu_{\tau}$
(if $ \sin^2\vartheta_{23} \neq 0.5 $).

However, the results of the analysis of the experimental data
with all the diagonal and transition charge radii
are complicated and suffer of degeneracies
in the effects of the different charge radii.
Most relevant is the possible cancellation of the contribution of a diagonal charge radius by the contribution of the corresponding transition charge radii.
This can be seen clearly in \cevns by noting that the main contribution to
$\left| Q_{\nu_{\ell}\mathcal{N}}^{\text{SM+CR}}(|\vet{q}|) \right|^2$
in Eq.\ref{eq:CRQW}
is due to $g_{V}^{n}$, which is negative (see Eq.\ref{eq:CEvNSgV}).
A small negative value of $\langle{r}_{\nu_{\ell}}^2\rangle$
which increases the positive contribution of
$ \left( \tilde{g}_{V}^{p} - \widetilde{Q}_{\ell\ell} \right) $,
leads to a decrease of the first part of
$\left| Q_{\nu_{\ell}\mathcal{N}}^{\text{SM+CR}}(|\vet{q}|) \right|^2$
which can be compensated by increasing the second part
with the transition charge radii.

Therefore, the analyses of the data without and with the
transition charge radii can give rather different results.
Moreover,
as we have seen in Eq.\ref{eq:SMchr},
the SM predicts the existence of only the diagonal flavor charge radii
and
it is plausible that the values of the transition charge radii
generated by BSM physics are much smaller.
Hence,
in Table~\ref{tab:CR}
we report the bounds
on the diagonal charge radii of the flavor neutrinos
obtained in different experiments and phenomenological analyses
assuming the absence of the transition charge radii.

From Table~\ref{tab:CR} one can see that the most stringent bounds on
$\langle r_{\nu_{e}}^2 \rangle^{\text{SM}}$
and
$\langle r_{\nu_{\mu}}^2 \rangle^{\text{SM}}$
\red{obtained in laboratory experiments}
are only about one order of magnitude larger than the
SM predictions in Eq.\ref{eq:SMchr},
\red{and the CCFR bound on
$\langle r_{\nu_{\mu}}^2 \rangle^{\text{SM}}$
obtained in Ref.\cite{Hirsch:2002uv} is near the SM prediction}.
Therefore,
an experimental effort is mandatory in order to try to improve
the experimental sensitivity and try to discover
the neutrino charge radii
predicted by the SM.

\red{In Table~\ref{tab:CR} we report also
limits on the specific charge radii obtained in solar \eves experiments,
generic limits on the charge radii obtained from searches of
$e^+ e^- \to \nu \bar\nu \gamma$
events in high-energy TRISTAN and LEP-2
experiments\footnote{Neglecting a possible $q^2$ dependence of the charge radius contribution to the cross section~\cite{AtzoriCorona:2024rtv}.},
and
generic limits on the charge radii obtained from
BBN, SN1987A, and TRGB.
Note that the BBN and SN1987A bounds are at the same level or smaller than the
SM predictions, but the uncertainties are unknown.}

The current bounds of solar neutrino experiments
on all the neutrino charge radii
are not competitive with the bounds obtained in laboratory experiments,
but we must take into account that
the results of solar low-energy \eves
are recent and come from experiments
which have been constructed for the search of dark matter,
in which solar neutrino interactions were considered as a background.
Future dark matter and solar neutrino experiments,
as
DARWIN~\cite{DARWIN:2020bnc,Giunti:2023yha},
XLZD~\cite{XLZD:2024gxx}, and
DarkSide~\cite{DarkSide-20k:2024yfq},
will reach higher sensitivities.

Accurate analyses of the data of future higher precision \eves and \cevns
experiments will require to take into account
the effects of the non-zero momentum transfer in the calculation of the
neutrino charge radius radiative correction~\cite{AtzoriCorona:2024rtv}.

\begin{table}[t!]
\hspace{-2cm}
\begin{minipage}{\textwidth}
\centering
\renewcommand{\arraystretch}{1.2}
\begin{tabular}{llcccc}
Method & Experiment & Limit $[e]$ & CL & Year & Ref.\\
\hline
\multirow{2}{*}{Neutrality of matter}
& K and Cs
& $\mathcal{Q}_{\nu} = ( - 0.5 \pm 3.5 ) \times 10^{-19}$
& 68\%
& 1988
& \cite{Hughes-Fraser-Carlson-1988}
\\[-0.35cm]
&
\multicolumn{5}{c}{\rule{0.8\linewidth}{0.5pt}}
\\[-0.15cm]
& $\text{SF}_{6}$~\cite{Bressi:2011yfa}
& $\mathcal{Q}_{\nu} = ( 0.6 \pm 3.2 ) \times 10^{-21}$
& 68\%
& 2014
& \cite{Giunti:2014ixa}~\footnote{Sign corrected.}
\\
\hline
Magnetic Dichroism
& PVLAS\footnote{For $m_{\nu} \lesssim 10 \, \text{meV}.$}
& $|Q_{\nu}| < 3 \times 10^{-8}$
& 95\%
& 2016
& \cite{DellaValle:2015xxa}
\\
\hline
\multirow{4}{*}{Reactor $\bar\nu_e$ E$\nu$ES}
& TEXONO~\cite{TEXONO:2002pra}
& $|Q_{\nu_{e}}|< 3.7 \times 10^{-12}$
& 90\%
& 2006
& \cite{Gninenko:2006fi}
\\
& GEMMA~\cite{Beda:2012zz}
& $|Q_{\nu_{e}}|< 1.5 \times 10^{-12}$
& 90\%
& 2013
& \cite{Studenikin:2013my}
\\
& TEXONO
& $|Q_{\nu_{e}}|< 1.0 \times 10^{-12}$
& 90\%
& 2014
& \cite{Chen:2014dsa}
\\
& CONUS
& $|Q_{\nu_{e}}|< 3.3 \times 10^{-12}$
& 90\%
& 2022
& \cite{CONUS:2022qbb}
\\
\hline
Accelerator $(\nu_{\mu},\bar\nu_{\mu})$ E$\nu$ES
& LSND~\cite{LSND:2001akn}
& $|Q_{\nu_{\mu}}|< 3 \times 10^{-9}$
& 90\%
& 2020
& \cite{Das:2020egb}
\\
\hline
Beam Dump $\nu_{\tau},\bar\nu_{\tau}$ E$\nu$ES
& BEBC~\cite{BEBCWA66:1986err}
& $|Q_{\nu_{\tau}}|< 4 \times 10^{-4}$
& 90\%
& 1993
& \cite{Babu:1993yh}
\\
\hline
Accelerator $\nu_{\tau},\bar\nu_{\tau}$ E$\nu$ES
& DONUT~\cite{DONUT:2001zvi}
& $|Q_{\nu_{\tau}}|< 4 \times 10^{-6}$
& 90\%
& 2020
& \cite{Das:2020egb}
\\
\hline
\multirow{2}{*}{\begin{tabular}{c}
Accelerator $\nu_{e},\nu_{\mu},\bar\nu_{\mu}$
\\[-0.1cm]
CE$\nu$NS+E$\nu$ES
\end{tabular}}
&
\begin{tabular}{c}
COHERENT
\\[-0.1cm]
\cite{COHERENT:2020iec,COHERENT:2021xmm}
\end{tabular}
&
$\begin{array}{c} \displaystyle
Q_{\nu_e} \in (-5.0,5.0) \times 10^{-10}
\\[-0.1cm] \displaystyle
Q_{\nu_\mu} \in (-1.9,1.9) \times 10^{-10}
\\[-0.1cm] \displaystyle
|Q_{\nu_{e\mu}}| < 1.8 \times 10^{-10}
\\[-0.1cm] \displaystyle
|Q_{\nu_{e\tau}}| < 5.0 \times 10^{-10}
\\[-0.1cm] \displaystyle
|Q_{\nu_{\mu\tau}}| < 1.9 \times 10^{-10}
\end{array}$
&90\%
&2022
& \cite{AtzoriCorona:2022qrf}
\\
\hline
\begin{tabular}{c}
Reactor $\bar\nu_{e}$
\\[-0.1cm]
CE$\nu$NS+E$\nu$ES
\end{tabular}
&
Dresden-II~\cite{Colaresi:2022obx}\footnote{Using the Fef quenching factor.}
&
$\begin{array}{c} \displaystyle
Q_{\nu_e} \in (-9.3,9.5) \times 10^{-12}
\\[-0.1cm] \displaystyle
|Q_{\nu_{e\mu}}|,|Q_{\nu_{e\tau}}| < 9.4 \times 10^{-12}
\end{array}$
&90\%
&2022
& \cite{AtzoriCorona:2022qrf}
\\
\hline
\multirow{8}{*}{Solar E$\nu$ES}
& XMASS-I
&
$\begin{array}{c} 
\displaystyle |Q_{\nu_{e}}| < 7.3 \times 10^{-12}
\\[-0.1cm]
\displaystyle |Q_{\nu_{\mu}}|,|Q_{\nu_{\tau}}| < 1.1 \times 10^{-11}
\end{array}$
& 90\%
& 2020
& \cite{XMASS:2020zke}
\\[-0.35cm]
&
\multicolumn{5}{c}{\rule{0.8\linewidth}{0.5pt}}
\\[-0.15cm]
& LUX-ZEPLIN~\cite{LZ:2022lsv}
&
$\begin{array}{c} 
\displaystyle Q_{\nu_{e}} \in (-2.1,2.0) \times 10^{-13}
\\[-0.1cm]
\displaystyle |Q_{\nu_{\mu}}|<3.1  \times 10^{-13}
\\[-0.1cm]
\displaystyle |Q_{\nu_{\tau}}| < 2.8 \times 10^{-13}
\end{array}$
& 90\%
& 2022
& \cite{AtzoriCorona:2022jeb}
\\[-0.35cm]
&
\multicolumn{5}{c}{\rule{0.8\linewidth}{0.5pt}}
\\[-0.15cm]
& XENONnT~\cite{XENON:2022ltv}
&
$\begin{array}{c} 
\displaystyle Q_{\nu_{e}} \in (-1.3,6.4) \times 10^{-13}
\\[-0.1cm]
\displaystyle Q_{\nu_{\mu}} \in (-6.2,6.1)  \times 10^{-13}
\\[-0.1cm]
\displaystyle Q_{\nu_{\tau}} \in (-5.4,5.2) \times 10^{-13}
\end{array}$
& 90\%
& 2022
& \cite{2208.06415}
\\[-0.35cm]
&
\multicolumn{5}{c}{\rule{0.8\linewidth}{0.5pt}}
\\[-0.15cm]
&
\hspace{-0.29cm}
\begin{tabular}{c}
LUX-ZEPLIN~\cite{LZ:2022lsv}
\\[-0.1cm]
$+$ PandaX-4T~\cite{PandaX:2022ood}
\\[-0.1cm]
$+$ XENONnT~\cite{XENON:2022ltv}
\end{tabular}
&
$\begin{array}{c} \displaystyle
Q_{\nu_{e}} \in (-2.0, 7.0) \times 10^{-13}
\\[-0.1cm] \displaystyle
Q_{\nu_{\mu}},Q_{\nu_{\tau}} \in (-7.5, 7.3) \times 10^{-13}
\end{array}$
& 90\%
& 2023
& \cite{Giunti:2023yha}
\\[-0.35cm]
&
\multicolumn{5}{c}{\rule{0.8\linewidth}{0.5pt}}
\\[-0.15cm]
& LUX-ZEPLIN
& $|Q_{\nu}| < 2.24 \times 10^{-13}$
& 90\%
& 2023
& \cite{LZ:2023poo}
\\
\hline
$\nu$ST 
& {}
& $|Q_{\nu}| \lesssim 1.3 \times 10^{-19}$
& {}
& 2012
& \cite{Studenikin:2012vi}
\\
\hline
SN1987A 
& {}
& $|Q_{\nu}| \lesssim 10^{-17}-10^{-15}$
& {}
& 1987
& \cite{Barbiellini:1987zz}
\\
\hline
\multirow{2}{*}{TRGB}
&
& $|Q_{\nu}| \lesssim 2 \times 10^{-14}$
&
& 1999
& \cite{Raffelt:1999tx}
\\[-0.35cm]
&
\multicolumn{5}{c}{\rule{0.8\linewidth}{0.5pt}}
\\[-0.15cm]
&
& $|Q_{\nu}| < 6.3 \times 10^{-15}$
& 95\%
& 2023
& \cite{Fung:2023euv}
\\
\hline
Solar Cooling
&
& $|Q_{\nu}| \lesssim 6 \times 10^{-14}$
&
& 1999
& \cite{Raffelt:1999gv}
\\
\hline
Magnetars
&
& $|Q_{\nu}| \lesssim 10^{-12}-10^{-11}$
&
& 2020
& \cite{Das:2020egb}
\\
\hline
\end{tabular}

\end{minipage}
\caption{ \label{tab:ech}
Experimental and astrophysical limits for the neutrino electric charges.
Here, $\mathcal{Q}_{\nu}$ is the diagonal charge of massive neutrinos
in Eq.\ref{eq:diagEC}
and
$Q_{\nu}$
denotes a generic neutrino charge or an appropriate effective neutrino charge.
}
\end{table}

\section{Neutrino electric charge}
\label{sec:EC}

Neutrinos are generally thought to be exactly neutral.
They are constrained to be exactly neutral in the SM
with only one generation
by the cancellation of the quantum axial triangle anomalies
(see the review in Ref.\citenum{Das:2020egb}),
which also implies the standard charge quantization of elementary particles.
Instead, in the three-generation SM
two neutrinos could have opposite charges,
and the third must be neutral~\cite{Foot:1990uf,Foot:1992ui}.
In this case the charges of the leptons of two generations are dequantized,
while the quark charges obey the standard charge quantization.
However,
this very exotic possibility is not considered in the definition of the SM.

Current BSM theories are constructed to give to neutrinos
the masses which are required by the observation of neutrino oscillations.
The constraints for the neutrino charges
depend on the Dirac or Majorana nature of neutrinos.
From Eq.\ref{ffmaj1},
massive Majorana neutrinos cannot have diagonal electric charges.
On the other hand, when right-handed neutrino fields are introduced in the theory,
the quantization of the charges of elementary particles is lost
and massive Dirac neutrinos can be charged
(see the review in Ref.\citenum{Das:2020egb}).
In general, massive Dirac neutrinos can have 
diagonal neutrino charges
$Q_{kk}$,
but they must be all equal,
because neutrinos are created as flavor neutrinos
($\nu_{e}$, $\nu_{\mu}$, or $\nu_{\tau}$)
which are superpositions of massive neutrinos.
Hence, charge conservation requires that
\begin{equation}
Q_{kk} = \mathcal{Q}_{\nu}
.
\label{eq:diagEC}
\end{equation}
For Majorana neutrinos $\mathcal{Q}_{\nu}=0$.

In scattering experiments,
since neutrino charges are probed in experiments with known neutrino flavors
the experimental bounds are given for the flavor charges~\cite{Kouzakov:2017hbc}
\begin{equation}
Q_{\nu_{\ell\ell'}}
=
\sum_{k,j}
U_{\ell k}
Q_{kj}
U_{\ell' j}^{*}
.
\label{echflavor}
\end{equation}
The hermiticity of the matrix of electric charges
is preserved by the unitary transformation \ref{echflavor}.
In the following,
for simplicity we denote
the real diagonal electric charges of the flavor neutrinos as
$Q_{\nu_{\ell}} \equiv Q_{\nu_{\ell\ell}}$.
Hence, we have
\begin{equation}
Q_{\nu_{\ell}}
=
\mathcal{Q}_{\nu}
+
2 \text{Re}
\sum_{k>j}
U_{\ell k}
Q_{kj}
U_{\ell j}^{*}
.
\label{eq:flavEC}
\end{equation}
Note that,
contrary to the equality of the diagonal charges in the mass basis,
the diagonal charges of Dirac neutrinos in the flavor basis can be different,
because of the mixing and the contributions
of the off-diagonal charges in the mass basis. The unequal diagonal charges of flavor neutrinos do not violate charge conservation, since flavor states are not physical states of neutrinos, as opposed to the mass states.
Also Majorana neutrinos can have diagonal charges in the flavor basis,
which are given by the contributions in Eq.\ref{eq:flavEC} of the off-diagonal charges in the mass basis if CP is violated by the Dirac or Majorana phases
of the mixing matrix
(from Eq.\ref{ffmaj1},
the off-diagonal charges of Majorana neutrinos are imaginary).
Since the trace of the neutrino electric charge matrix is invariant under the unitary transformation \ref{echflavor}, the sum of the diagonal electric charges of flavor neutrinos (of both Dirac and Majorana types) is subject to the condition
\begin{equation}
\sum_{\ell}Q_{\nu_{\ell}}
=
\sum_{k}Q_{kk}=3\mathcal{Q}_{\nu}
,
\label{echflavor_diag}
\end{equation}
with $\mathcal{Q}_{\nu}=0$ for Majorana neutrinos.
Therefore, in the case of Majorana neutrinos,
the charges of the flavor neutrinos
cannot have the same sign.

The very stringent bounds in Table~\ref{tab:ech}
which have been obtained from the measurements of the neutrality of matter
have been inferred from charge conservation in neutron $\beta$ decay
(see, e.g., Ref.\citenum{Giunti:2014ixa}).
Therefore,
they constrain only the diagonal common charge
$\mathcal{Q}_{\nu}$
of the massive neutrinos which constitute the $\bar\nu_{e}$
produced in neutron $\beta$ decay
and the bound is non-trivial only
for Dirac neutrinos
(for Majorana neutrinos $\mathcal{Q}_{\nu}=0$).
The other bounds in Table~\ref{tab:ech}
are less stringent, but they are essential to get information
on the contributions of the off-diagonal charges of massive neutrinos
to the diagonal and off-diagonal charges of flavor neutrinos.

Considering neutrino scattering experiments,
note that the charge term in the vertex function conserves the helicity
of ultrarelativistic neutrinos
(see, e.g, Ref.\citenum{Giunti:2014ixa}),
as the weak neutral current SM interaction.
Taking into account that the diagonal (off-diagonal) charges in the flavor basis 
induce flavor-conserving (flavor-changing) interactions,
as weak neutral current SM interactions,
in each scattering process of a flavor neutrino
the contribution of the corresponding diagonal (off-diagonal) charge
add coherently (incoherently) to the weak neutral current. Specifically, for elastic scattering of a flavor neutrino on a free electron (\eves) or a nucleus (\cevns) the effect of $Q_{\nu_{\ell}}$ is the shift of the corresponding vector-coupling coefficient:
$$
g_V^{\nu_\ell} \to g_V^{\nu_\ell}-\frac{2m_W^2}{m_eT_e}\left(\frac{Q_{\nu_{\ell}}}{e}\right)\sin^2\vartheta_W,  \qquad g_V^p(\nu_\ell) \to g_V^p(\nu_\ell)+\frac{2m_W^2}{M_\mathcal{N}T_\mathcal{N}}\left(\frac{Q_{\nu_{\ell}}}{e}\right)\sin^2\vartheta_W.
$$
At low values of $T=T_{e(\mathcal{N})}$ the purely electric-charge part of the cross section, $d\sigma(Q_{\nu_{\ell}})/dT\propto Q_{\nu_{\ell}}^2$, has an approximate $1/T^2$ dependence. This can be used to constrain $Q_{\nu_{\ell}}$ from the upper bound $\mu_{\nu_{\ell}}^{\rm up}$ on $\mu_{\nu_{\ell}}$ by demanding that the ratio of the $d\sigma(Q_{\nu_{\ell}})/dT$ and $d\sigma(\mu_{\nu_{\ell}})/dT$ cross sections at the lower threshold recoil energy $T^{\rm th}$ measured in the detector does not exceed one. Following this approach, originally proposed in Ref.\cite{Studenikin:2013my}, one obtains in the \eves case
\begin{equation}
\label{ech:upbound}
    |Q_{\nu_\ell}|\lesssim\sqrt{\frac{T^{\rm th}_e}{2m_e}}\left(\frac{\mu_{\nu_\ell}^{\rm up}}{\mu_{\text{B}}}\right)e.
\end{equation}
The \eves experimental limits, except the BEBC~\cite{Babu:1993yh}, TEXONO~\cite{Chen:2014dsa}, LUX-ZEPLIN~\cite{AtzoriCorona:2022jeb} and XENONnT~\cite{2208.06415} limits, presented in Table~\ref{tab:ech} are obtained on the basis of Eq.\ref{ech:upbound}. The BEBC limit for tau neutrinos was derived employing the ratio of the integrated cross sections $\sigma(Q_{\nu_\tau})/\sigma(\mu_{\nu_\tau})$ corresponding to the electron recoil energies $T_e\geq T^{\rm th}_e$ rather than the differential ones as in Eq.\ref{ech:upbound}. The TEXONO limit~\cite{Chen:2014dsa} is obtained independently of the $\mu_\nu$ limit. It uses full spectral data in the statistical analysis and takes into consideration the fact that when $T^{\rm th}_e$ is close to the atomic ionization energy the cross section $d\sigma(Q_{\nu_{\ell}})/dT_e$ can strongly deviate from that in the free-electron approximation due to electron binding in the atoms of the detector material. Qualitatively this can be explained using the equivalent photon approximation (EPA) for scattering of a relativistic charged lepton on a target. It yields the differential cross section of the atomic ionization process induced by the neutrino electric charge in the form (see also Ref.\citenum{Chen:2014dsa})
\begin{equation}
\label{ech:EPAcrsec}
    \left(\frac{d\sigma(Q_{\nu})}{dT_e}\right)_{\rm EPA}=\frac{2\alpha}{\pi}\left(\frac{Q_{\nu}}{e}\right)^2\left(\frac{\sigma_\gamma(T_e)}{T_e}\right)\log\frac{E_\nu}{m_\nu},
\end{equation}
where $\sigma_\gamma(T_e)$ is the photoionization cross section for the photon energy $E_\gamma=T_e$. At low $T_e$, the cross section $\sigma_\gamma(T_e)$ has an approximate $1/T_e^3$ dependence, which leads to an enhanced sensitivity to the neutrino electric charge than is assumed using the free-electron approximation. It should be noted that the recent LUX-ZEPLIN bounds~\cite{AtzoriCorona:2022jeb} for solar neutrinos in Table~\ref{tab:ech} are derived employing the equivalent photon approximation~\ref{ech:EPAcrsec}, whereas the XENONnT bounds~\cite{2208.06415} correspond to the free-electron approximation. A bit weaker bounds based on the LUX-ZEPLIN and XENONnT data have been recently obtained in Ref.\cite{Giunti:2023yha} using the free-electron approximation and more conservative treatment of systematic uncertainties in the analysis. 

Similar to the case of the neutrino magnetic moments,
\eves experiments are able to probe smaller values of the neutrino electric charges than \cevns experiments,
as can be seen from Table~\ref{tab:ech}.
However, the most stringent constraint\footnote{Although
it is considered as the dominant bound by the
Particle Data Group~\cite{ParticleDataGroup:2024cfk},
we do not consider the controversial~\cite{Karshenboim:2024iff}
very stringent bound
$ |Q_{\nu}| \lesssim 4 \times 10^{-35} \, e $
which was obtained in Ref.\cite{Caprini:2003gz}
from a limit on the charge asymmetry of the Universe
under several unproved assumptions,
as the crucial assumption that there are no cancellations
between the charge asymmetries of different particle species
(as the cancellation between neutrinos and antineutrinos).}
on a generic neutrino charge $Q_{\nu}$
is the astrophysical bound
obtained in Ref.\cite{Studenikin:2012vi} by considering the impact of the
``neutrino star turning'' mechanism ($\nu$ST) that accounts for the motion of a millicharged neutrino along a curved trajectory inside a magnetized star,
which can shift the rotation frequency of the magnetized pulsar.
The millicharged neutrinos escaping from the star follow
a trajectory which is not perpendicular to the surface of the star.
This affects the initial star rotation during the formation of the pulsar in a supernova explosion.
The relative frequency shift of a born pulsar due to the $\nu$ST mechanism is given by
\begin{equation}
\label{delta_omega_nonzero_charge}
\frac{|\triangle\omega|}{\omega_0}=7.6\left(\frac{Q_{\nu}}{e}\right)\times
10^{18}\left(\frac{P_0}{10\text{ s}}\right)
\left(\frac{N_{\nu}}{10^{58}}\right)
\left(\frac{1.4  M_{\odot}}{M_{S}}\right)
\left(\frac{B}{10^{14}\text{ G}}\right),
\end{equation}
where $P_0= 10$~s is a pulsar initial spin period and $N_{\nu} \sim 10^{58}$ is the number of emitted neutrinos. The $\nu$ST limit on the neutrino millicharge in Table~\ref{tab:ech} follows from the straightforward demand $|\triangle\omega| < \omega_0$ and Eq.\ref{delta_omega_nonzero_charge}. 

In Ref.\cite{Barbiellini:1987zz} the limit on the neutrino millicharge was obtained from a different approach based on the consideration of the millicharged neutrino flux from SN1987A propagating in the galactic and intergalactic magnetic fields outside the star. For larger values of the electric charge of neutrinos the magnetic fields would lengthen their path, and neutrinos of different energy could not arrive on the Earth within a few seconds of each other, even if emitted simultaneously by the supernova. 

\section{Astrophysical neutrino electromagnetic processes}
\label{sec:Astrophysical}

The couplings with real and virtual photons of
neutrinos with nonzero electromagnetic properties
generate processes that can occur in various astrophysical conditions
and can be the cause of important observable phenomena. 

\begin{figure}
\begin{minipage}{0.3\textwidth}
\includegraphics[clip, width=\textwidth, viewport=22 239 366 454]{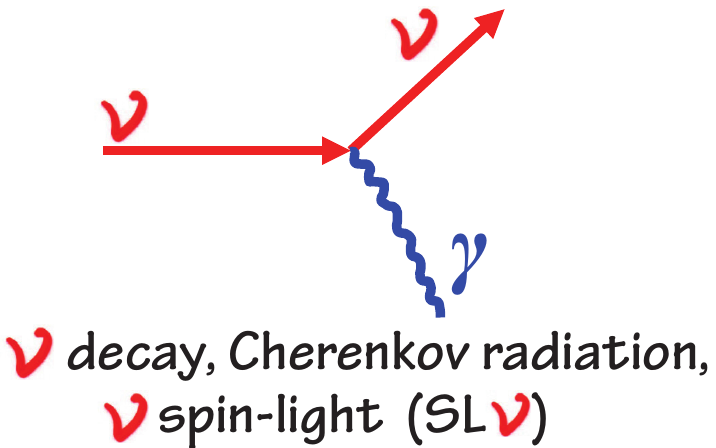}
\end{minipage}
\hfill
\begin{minipage}{0.2\textwidth}
\includegraphics[clip, width=\textwidth, viewport=417 263 634 453]{figures/new_figure_nu_em_inter.pdf}
\end{minipage}
\hfill
\begin{minipage}{0.23\textwidth}
\includegraphics[clip, width=\textwidth, viewport=50 28 307 219]{figures/new_figure_nu_em_inter.pdf}
\end{minipage}
\hfill
\begin{minipage}{0.23\textwidth}
\includegraphics[clip, width=\textwidth, viewport=403 22 675 223]{figures/new_figure_nu_em_inter.pdf}
\end{minipage}
\caption{Astrophysical neutrino electromagnetic processes.}
\label{fig:astro}
\end{figure}

The most important neutrino electromagnetic processes are shown in Fig.\ref{fig:astro}
(see, e.g., Refs.\citenum{Giunti:2014ixa,Studenikin:2022rhv,Raffelt:1996wa}).
They are the following: 1) a heavier neutrino decay to a lighter mass state in vacuum, 2) the Cherenkov radiation by a neutrino in matter or an external magnetic field, 3) the spin-light of neutrino in matter, 4) the plasmon decay to a neutrino-antineutrino pair in matter, 5) the neutrino scattering on an electron or a nucleus, and 6) the neutrino spin precession in an external magnetic field and transversely moving (or transversely polarized) matter.
All of these processes can be of great interest in astrophysics.
The registration of the possible consequences of these processes allows us to obtain information about the values of the electromagnetic properties of neutrinos
and also to set appropriate limits.
So, astrophysics can be considered as a laboratory for studying the electromagnetic properties of neutrinos (see, e.g., Refs.\citenum{Giunti:2014ixa,Raffelt:1999tx,Brdar:2020quo}).

\subsection{Neutrino radiative decay}
\label{sec:decay}

If neutrinos have nonzero electric charges or
transition magnetic or electric moments,
a neutrino mass state $\nu_i$ can decay into a lighter state $\nu_f$
with emission of a photon.
This process was discussed for the first time in Ref.\cite{Shrock:1974nd} and
concrete applications to neutrinos can be found in Refs.\cite{Petcov:1976ff,Zatsepin:1978iy}.
For more recent papers and a detailed discussion of neutrino radiative decay
$\nu_i \rightarrow \nu_f + \gamma$, see Refs.\cite{Giunti:2014ixa,Balaji:2019fxd,Balaji:2020oig}.
If one neglects the effects of the neutrino charges and charge radii,
the neutrino electromagnetic vertex function \ref{eq:Lambda2} reduces to
\begin{equation}
\Lambda^{\mu}_{fi}(q)
=
-
i \sigma^{\mu\nu} q_{\nu}
\mu_{fi}
,
\label{eq:Lambda2decay}
\end{equation}
where, as before,
$\mu_{fi}$ is an effective magnetic moment which includes possible electric moment
contributions and
we dropped the ``eff'' superscript for simplicity.

The decay rate in the rest frame of the decaying neutrino $\nu_{i}$
is given by~\cite{Marciano:1977wx,Lee:1977tib}
\begin{equation}
\Gamma_{\nu_{i}\to\nu_{f}+\gamma}
= \frac{1}{8\pi} \left( \frac{m_{i}^{2}-m_{f}^{2}}{m_{i}}
\right)^3 |\mu_{fi}|^{2}
\simeq
5.4
\left( \dfrac{m^{2}_{i}-m^{2}_{f}}{m_{i}^{2}} \right)^3
\left( \dfrac{m_i}{1 \, \text{eV}} \right)^3
\left| \dfrac{\mu_{fi}}{\mu_{\text{B}}} \right|^2
\text{s}^{-1}
.
\label{eq:decrat}
\end{equation}
Note that since $m_i \neq m_f$ only the transition magnetic and electric moments contribute.
Therefore, Eq.\ref{eq:decrat} is equally valid  for both Dirac and Majorana neutrinos.

In the simplest extensions of the SM
the Dirac and Majorana transition magnetic moments
are smaller than $10^{-23} \, \mu_{\text{B}}$
(see Eqs.\ref{muDtran} and~\ref{muDtranbf},
and Ref.\citenum{Shrock:1982sc})
and the life time is huge:
\begin{equation}
\tau_{\nu_i\rightarrow \nu_j +\gamma}
\simeq
0.19 \times 10^{46}
\left( \dfrac{m^2_i}{m^2_i-m^2_j} \right)^3
\left( \dfrac{1\,\text{eV}}{m_i} \right)^3
\left| \dfrac{10^{-23}\mu_{\text{B}}}{\mu_{fi}} \right|^2
\text{s}
.
\end{equation}

The neutrino radiative decay has been constrained from the absence of decay photons in studies of the solar, supernova and reactor (anti)neutrino fluxes, as well as from the absence of the spectral distortions of the cosmic microwave background radiation. However, the corresponding upper bounds on the effective neutrino magnetic moments \cite{Raffelt:1999tx} are in general less stringent than the astrophysical bounds from plasmon decay into a neutrino-antineutrino pair discussed in Section~\ref{sec:Plasmon}.

\subsection{Neutrino Cherenkov radiation and related processes}
\label{sec:Ch_rad}

Neutrinos with a magnetic moment
(and/or an electric charge, as well as an electric moment)
propagating in a medium with a velocity larger than the velocity of light in the medium can emit Cherenkov radiation.
This possibility was considered for the first time in Ref.\cite{Radomski:1975re}.

The rate of Cherenkov radiation has been calculated in Ref.\cite{Grimus:1993np}.
Considering solar neutrinos, the authors of Ref.\cite{Grimus:1993np}
found that with an effective magnetic moment of
$10^{-11} \mu_{\text{B}}$
about five Cherenkov photons
in the range of visible light would be emitted per day in a
$\text{km}^3$ water detector.  

In the presence of a magnetic field and/or matter, neutrinos can emit photons by specific mechanisms due to the induced $\nu - \gamma$ effective vertex and the modified photon dispersion relation.
In these environments neutrinos can acquire an induced magnetic moment and also an induced electric charge, so that light can be emitted even in the case of massless neutrinos and without the need of BSM effects
(see Refs.\citenum{Galtsov:1972xp,Giunti:1992sy,Oraevsky:1986dt,Ioannisian:1996pn,Ternov:2016njz,Giunti:2014ixa}). 

\subsection{Plasmon decay into a neutrino-antineutrino pair}
\label{sec:Plasmon}

The most interesting astrophysical process
for constraining neutrino electromagnetic properties,
in particular neutrino magnetic moments,
is plasmon decay into a neutrino-antineutrino pair~\cite{Bernstein:1963qh}.
This process becomes kinematically allowed in media where a photon behaves as a particle with an effective mass $\omega_{\text{P}}$.
In a nonrelativistic plasma the dispersion relation for a photon (plasmon) is
$\omega_{\gamma}^2-{\vec k}_{\gamma}^2=\omega_{\text{P}}^2$,
where $\omega_{\text{P}} = \sqrt{ 4 \pi \alpha N_e / m_e }$
is the plasma frequency
($N_e$ is the electron density)~\cite{Raffelt:1996wa}.
The plasmon decay rate is given by~\cite{Raffelt:1996wa}
\begin{equation}
\Gamma_{\gamma^{*}\to\nu{\bar\nu}}
(\mu_{\nu})
=
\frac{\mu_{\nu}^{2}}{24\pi}
\,
Z
\,
\frac{\omega_{\text{P}}^4}{\omega_\gamma}
,
\label{eq:plasmondecratMM}
\end{equation}
where $Z$ is a factor which depends on the polarization of the plasmon
and the effective magnetic moment is given by
\begin{equation}
\mu_{\nu}^2
=
\sum_{k,j}
\left(
|\mu_{kj}|^{2}
+
|\epsilon_{kj}|^{2}
\right)
.
\label{eq:mueff-plasmon}
\end{equation}
A plasmon decay into a neutrino-antineutrino pair transfers the energy $\omega_\gamma$ to neutrinos that can freely escape from a star and thus can increase the star cooling speed.
The corresponding energy-loss rate per unit volume is
\begin{equation}
Q_{\gamma^{*}\to\nu {\bar\nu}}=\frac{g}{(2
\pi)^3}\int\Gamma_{\gamma^{*} \to\nu {\bar
\nu}}f_{k_{\gamma}}\omega_{\gamma}d^3k_{\gamma},
\end{equation}
where $f_{k_\gamma}$ is the photon Bose-Einstein distribution function and $g=2$ is the number of polarization states.
This energy loss can delay the helium ignition in low-mass red giants
and increase the
Tip of the Red Giant Branch (TRGB)
brightness~\cite{Raffelt:1996wa},
leading to the TRGB bounds in Table~\ref{tab:MM-AST}.
Also the Solar Cooling, Cepheid Stars, and White Dwarfs bounds
in Table~\ref{tab:MM-AST}
follow from the energy loss induced by plasmon decay.

The plasmon decay can also be used to constrain
the neutrino electric charges~\cite{Raffelt:1999tx}.
The rate of plasmon decay to a neutrino-antineutrino pair due to the effective
neutrino charge
$Q_{\nu}=\sqrt{\sum_{k,j}|Q_{kj}|^2}$
is given by~\cite{Raffelt:1996wa}
\begin{equation}
\Gamma_{\gamma^{*}\to\nu{\bar\nu}}
(Q_{\nu})
=
\frac{\alpha Q_{\nu}^{2}}{3}
\,
Z
\,
\frac{\omega_{\text{P}}^2}{\omega_\gamma}
.
\label{eq:plasmondecratEC}
\end{equation}
The corresponding TRGB and solar cooling constraints on $Q_{\nu}$ are listed in Table~\ref{tab:ech}, along with somewhat weaker constraints that are obtained taking into account the Schwinger production of charged $\nu\bar\nu$ pairs in the strong electromagnetic field in magnetars.

\subsection{Spin light of neutrino}
\label{sec:Spin_light}

It is known from
classical electrodynamics that a system with zero electric charge but nonzero magnetic (or electric) moment
can produce electromagnetic radiation which is called ``magnetic (or electric) dipole radiation''.
It is due to the rotation of the magnetic (or electric) moment. 
A similar mechanism of radiation exists
in the case of a neutrino
with a magnetic (or electric) moment propagating in matter
\cite{Lobanov:2002ur}.
This phenomenon,
called ``spin light of neutrino''
(SL$\nu$),
is different from
the neutrino Cherenkov radiation in matter discussed in Subsection~\ref{sec:Ch_rad},
because it can exist even when the emitted photon refractive index
is equal to unity.
The SL$\nu$ is a radiation produced by the neutrino on its own,
rather than a radiation of the background particles.
Since the SL$\nu$ process is a transition between neutrino states with equal masses,
it can only become possible because of an external environment
influence on the neutrino states.

The SL$\nu$ was first studied
with a quasi-classical treatment based on a Lorentz-invariant approach to the neutrino spin
evolution that implies the use of the generalized Bargmann-Michel-Telegdi equation. The full quantum theory of the SL$\nu$ has been elaborated in Ref.\cite{Studenikin:2004dx}.
The quantum theory of the SL$\nu$ method is based on the exact solutions of the modified Dirac equation for the neutrino wave function in matter. 

The Feynman diagram of the SL$\nu$ process is shown in
Fig.\ref{fig:astro},
where the neutrino initial and final states
are exact solutions of the corresponding
Dirac equations accounting for the interactions with matter.
Here we consider a generic flavor neutrino with an effective magnetic moment $\mu_{\nu}$ and effective mass $m_{\nu}$.
The SL$\nu$ process for a relativistic neutrino
is a transition from a more energetic neutrino initial state to a
less energetic final state with emission of a photon
and a neutrino helicity flip.

The amplitude of the SL$\nu$ process is given by 
\begin{equation}
S_{fi}
=
- \mu_{\nu} \sqrt{4\pi}
\int d^{4} x
{\bar\psi}_{f}(x)(\vec\Gamma \cdot \vec\varepsilon^{*})\frac{e^{ikx}}{\sqrt{2\omega L^{3}}}\psi_{i}(x)
,
\quad
\text{where}
\quad
\vec{\Gamma}=i\omega\big\{\big[\vec{\Sigma} \times
\vec{\varkappa}\big]+i\gamma^{5}\vec{\Sigma}\big\}
,
\quad
\vec\Sigma
=
\begin{pmatrix}
\vec\sigma & 0
\\
0 & \vec\sigma
\end{pmatrix}
,
\label{amplitudeGamma}
\end{equation}
and $k^{\mu}=(\omega,{\vec k})$ and $\vec\varepsilon$ are the photon momentum and polarization
vectors, $\vec{\varkappa}=\vec{k}/{\omega}$ is the unit vector
pointing in the direction of propagation of the emitted photon.
Here $\psi_{i}(x)$ and $\psi_{f}(x)$ are the
initial and final neutrino wave functions in presence of matter
obtained as exact solutions of the effective Dirac equation. 
For a relativistic neutrino with momentum $p \gg m_{\nu}$,
the total rate $\Gamma$ and power $I$
of SL$\nu$ for different ranges of the effective matter density $\widetilde{N}$ are
\begin{equation}\label{pgg}
\Gamma = \left\{
\begin{array}{lcl}
\frac{64}{3} \mu_{\nu}^{2} \widetilde{N}^3 p^2 m_{\nu}
&\text{for}&
\widetilde{N} \ll \frac{m_{\nu}}{p},
\\
4 \mu_{\nu}^{2} \widetilde{N}^2 m_{\nu}^2 p
&\text{for}&
\frac{m_{\nu}}{p} \ll \widetilde{N} \ll \frac{p}{m_{\nu}},
\\
4 \mu_{\nu}^{2} \widetilde{N}^3 m_{\nu}^3
&\text{for}&
\widetilde{N} \gg \frac{p}{m_{\nu}},
\end{array}
\right.
\qquad
\qquad
I = \left\{
\begin{array}{lcl}
\frac{128}{3} \mu_{\nu}^{2}\widetilde{N}^{4}p^{4}
&\text{for}&
\widetilde{N} \ll\frac{m_{\nu}}{p},
\\
\frac{4}{3} \mu_{\nu}^{2} \widetilde{N}^2 m_{\nu}^2 p^2
&\text{for}&
\frac{m_{\nu}}{p} \ll \widetilde{N} \ll \frac{p}{m_{\nu}},
\\
4 \mu_{\nu}^{2} \widetilde{N}^4 m_{\nu}^4
&\text{for}&
\widetilde{N} \gg \frac{p}{m_{\nu}}.
\end{array}
\right.
\end{equation}
For a $\nu_{e}$ moving in a medium composed of electrons, protons, and neutrons,
the effective matter density is given by
\begin{equation}
\widetilde{N} = \frac{G_F}{2{\sqrt 2}}
\Big[N_{e} (1+4\sin^2\theta_W )+ N_{p}(1-4\sin^2\theta_W )-N_{n}\Big]
,
\label{Ntilde}
\end{equation}
where $N_{e}, N_{p}$, and $N_{n}$ are the number densities of the background electrons, protons, and neutrons, respectively.

Some specific features of this process might have
phenomenological consequences in astrophysics
(see Ref.\citenum{Grigoriev:2017wff} for a detailed discussion).
For a wide range of matter densities,
the SL$\nu$ rate and power increase with the neutrino momentum.
For ultrahigh-energy neutrinos ($p \sim 10^{18} eV $)
propagating through a dense medium with
$\widetilde{N} \sim 10 \, \text{eV}$
(this value is typical for a neutron star with
$N_{n} \sim 10^{38}\,\text{cm}^{-3}$),
the rate of the SL$\nu$ process is about $0.7\,\text{s}^{-1}$.
This is the most favorable case
for the manifestation of the effects
of SL$\nu$.
It can be realized in galaxy clusters~\cite{Grigoriev:2017wff}.
Moreover,
the SL$\nu$ polarization properties
may be related to the observed polarization of
gamma ray bursts (GRB)~\cite{Grigoriev:2017wff}.

\subsection{Neutrino spin and spin-flavor oscillations in magnetic fields and moving matter}
\label{sec:spin_oscillations}

The neutrino spin oscillations $\nu_{L} \leftrightarrow \nu_{R}$ induced by the neutrino magnetic moment interaction with a transversal magnetic field $\vec{B}_{\perp}$ were first considered in Ref.\cite{Cisneros:1970nq}.
Then spin-flavor oscillations $\nu_{e L} \leftrightarrow \nu_{\mu R}$ in $\vec{B}_{\perp}$ in vacuum were discussed
in Ref.\cite{Schechter:1981hw} and
the importance of the matter effect was emphasized in Ref.\cite{Okun:1986hi}.
The effect of the resonant amplification of neutrino spin oscillations in $\vec{B}_{\perp}$ in the presence of matter was proposed in Refs.\cite{Akhmedov:1988uk,Lim:1987tk}.
Neutrino spin oscillations in a magnetic field with
account for the effect of moving matter was studied in Ref.\cite{Lobanov:2001ar}.
A possibility to establish
conditions for the resonance in neutrino spin oscillations by the effect of matter motion was discussed in
Ref.\cite{Studenikin:2004bu}.
The impact of the longitudinal magnetic field $\vec{B}_{||}$ was discussed in Ref.\cite{Akhmedov:1988hd}. The effects of non-zero Dirac and Majorana CP violating phases on neutrino-antineutrino oscillations in a magnetic field of astrophysical environments was investigated in Ref.\cite{Popov:2021prd}.  In particular, it was shown that neutrino-antineutrino oscillations combined with Majorana-type CP violation can affect the $\bar\nu_e/\nu_e$ ratio for neutrinos coming from the supernovae explosion. 

The neutrino spin oscillations in the presence of a constant twisting magnetic field
were considered in Refs.\cite{Vidal:1990fr,Smirnov:1991ia,Akhmedov:1993sh,Likhachev:1990ki,Dvornikov:2007aj,Dmitriev:2015ega}.
In particular, in Ref.\cite{Likhachev:1990ki} the possibility of the transition of up to half of the active left-handed neutrinos into sterile right-handed neutrinos when the neutrino flux leaves the surface of a dense magnetized neutron star (the ``cross-border effect'') was predicted.

The resonant spin conversion of neutrinos induced by the geometrical phase in a twisting magnetic field has been studied
in Ref.\cite{Jana:2023ufy}.
It has been shown that 
it could affect the supernova neutronization bursts where very intense magnetic fields are likely.
Assuming this mechanism,
the measurement of the flavor composition of supernova neutrinos
in upcoming neutrino experiments like DUNE and Hyper-Kamiokande
can be used as a probe of neutrino magnetic moments, potentially down to a few $10^{-15} \mu_{\text{B}}$~\cite{Jana:2023ufy}.

In Ref.\cite{Egorov:1999ah} neutrino spin oscillations were considered in the presence of an arbitrary constant electromagnetic field $F_{\mu\nu}$. Neutrino spin oscillations in the presence of the field of circular and linearly polarized electromagnetic waves and superposition of an electromagnetic wave and a constant magnetic field were considered in Refs.\cite{Lobanov:2001ar,Dvornikov:2001ez}.
The effect of the parametric resonance in neutrino oscillations in periodically varying electromagnetic fields
was studied in Ref.\cite{Dvornikov:2004en}.

The more general case of neutrino spin evolution for a neutrino subjected to general types of non-derivative interactions with external scalar $s$, pseudoscalar $\pi$, vector $V_{\mu}$, axial-vector $A_{\mu}$, tensor $T_{\mu\nu}$ and pseudotensor $\Pi_{\mu\nu}$ fields was considered in
Ref.\cite{Dvornikov:2002rs}. From the  general neutrino spin evolution equation, obtained in Ref.\cite{Dvornikov:2002rs}, it follows that neither scalar $s$,
nor pseudoscalar $\pi$, nor vector $V_{\mu}$ fields can induce a neutrino spin evolution. On the contrary, it was shown that electromagnetic (tensor) and weak (axial-vector) interactions can contribute to the neutrino spin evolution.

Neutrino mixing and oscillations in an arbitrary constant magnetic field that have nonzero $\vec{B}_{\perp}$ and $\vec{B}_{||}$ components were considered in Refs.\cite{Fabbricatore:2016nec,Studenikin:2016oyh,Kurashvili:2017zab}. 

\section{Neutrino magnetic moment portal}
\label{sec:portal}

Singlet neutral fermions of the Standard Model (SM) gauge group are often introduced to account for the non-zero neutrino mass and are referred to as sterile neutrinos. These sterile neutrinos can interact with the SM sector either through the active-to-sterile mixing portal (see the review in Ref.\citenum{Dasgupta:2021ies}) or via the active-to-sterile dipole moment portal. The neutrino dipole portal (NDP) is characterized by the dimension-five operator~\cite{Schwetz:2020xra,Magill:2018jla}
\begin{equation}
	\mathcal{L}_{\rm NDP}=
	\bar{N}(i\slashed{\partial}-m_N)N+\frac{1}{2}d_\ell\bar{N}\sigma_{\mu\nu}\nu_\ell F^{\mu\nu}
	+ \text{h.c.}
	,
	\label{lagrangian}	
\end{equation}
where $\nu_\ell$ is the active neutrino of flavor $\ell$, $N$ is the sterile neutrino with a mass of $m_N$, and $d_\ell$ is the neutrino transition magnetic moment.
In the literature, the unit of $d_\ell$ is often chosen as
$\text{GeV}^{-1}$, which is inversely proportional to the scale of new physics and can be connected with the traditional unit of $\mu_{\text{B}}$ by the relation
\begin{equation}
    d_\ell=\frac{\sqrt{\pi\alpha}}{m_e}\left| \frac{\mu_{\nu_\ell}}{\mu_{\text{B}}}\right|\simeq 296\,{\text{GeV}}^{-1}
    \left| \frac{\mu_{\nu_\ell}}{\mu_{\text{B}}} \right|
    .
\end{equation}

For a broad range of sterile neutrino masses $m_{N}$,
the magnetic moment $d_\ell$ can be investigated using a variety of methods. In the high mass regime, extending up to 100 $\text{GeV}$, measurements at the electron-positron collider experiments, such as LEP, are capable of probing the magnetic moment at the level of 
$d_\ell\sim 10^{-3}\,\text{GeV}^{-1}$~\cite{Magill:2018jla}. In the sub-GeV mass range, sterile neutrinos produced in core-collapse supernovae may escape, carrying away additional energy. This energy loss mechanism imposes constraints, excluding the range 
$ 10^{-7} \, \text{GeV}^{-1} \lesssim d_\ell \lesssim 10^{-3} \, \text{GeV}^{-1}$~\cite{Magill:2018jla}.
Below this exclusion band, searches for decay products of sterile neutrinos originating from supernovae,
conducted at neutrino detectors~\cite{Brdar:2023tmi,Chauhan:2024nfa,Lazar:2024ovc} and $\gamma-$ray telescopes~\cite{Brdar:2023tmi},
can potentially probe even smaller values of $d_\ell$.
To investigate the regions above the supernova cooling band, we can employ high intensity beam dump and neutrino experiments.
There are searches via upscattering of atmospheric and solar neutrinos to $N$ which decays at detectors~\cite{Plestid:2020vqf,Gustafson:2022rsz}, upscattering at Forward LHC detectors~\cite{Ismail:2021dyp},
Borexino and CHARM-II scattering measurements~\cite{Brdar:2020quo,BOREXINO:2018ohr,Coloma:2017ppo,CHARM-II:1991ydz},
and the spectrum distortion of neutrino scattering at dark matter detectors~\cite{Shoemaker:2018vii,Brdar:2020quo}.
With all these data, $d_{\ell}$ regions above the supernova cooling band for $m_N \lesssim 10 \,\text{MeV}$ can be probed.
For larger $m_N$ up to $50\,\text{MeV}$,
existing studies can cover the region $d_{\ell} \gtrsim 5 \times 10^{-7} \, \text{GeV}^{-1}$,
leaving an unexplored region for smaller $d_{\ell}$.
The \cevns data of reactor neutrinos, neutrinos from the Spallation Neutron Source,
and solar neutrinos can address this gap,
offering competitive and complementary constraints on sterile neutrino masses up to about 50 MeV.

For the \cevns process induced by the NDP when a $\nu_{\ell}$ interacts with a nucleus ${}^{A}_{Z}\mathcal{N}$, the cross section is
\begin{equation}
\dfrac{d\sigma^{\rm NDP}_{\nu_{\ell}}}{d T_{\mathcal{N}}}=d_{\ell}^2 \alpha Z^{2}
[F_{Z}^{\mathcal{N}}(|\vet{q}|)]^2
\left[
\frac{1}{T_{\mathcal{N}}}-\frac{1}{E_{\nu}}-\frac{m_{N}^2}{2 E_{\nu} M_{\mathcal{N}}T_{\mathcal{N}}}\left(1-\frac{M_{\mathcal{N}}-T_{\mathcal{N}}}{2E_{\nu}}\right)-\frac{m_{N}^4(M_{\mathcal{N}}-T_{\mathcal{N}})}{8E_{\nu}^{2}M_{\mathcal{N}}^2T_{\mathcal{N}}^2}
\right]
.
\label{crosssection:NDP} 
\end{equation}
Since the sterile neutrino mass can vary widely, the third term becomes significant when $m_{N}$ is comparable to the energy of the incident neutrino.
This offers a unique opportunity to probe the neutrino dipole portal for values of $m_{N}$ in this specific energy range.
In the following we are going to show that the \cevns interactions of neutrinos from reactors, Spallation Neutron Source, and the Sun can provide unique probes of $d_{e}$, $d_{\mu}$, and $d_{\tau}$.

\begin{figure}
		\centering
		\includegraphics[scale=0.16]{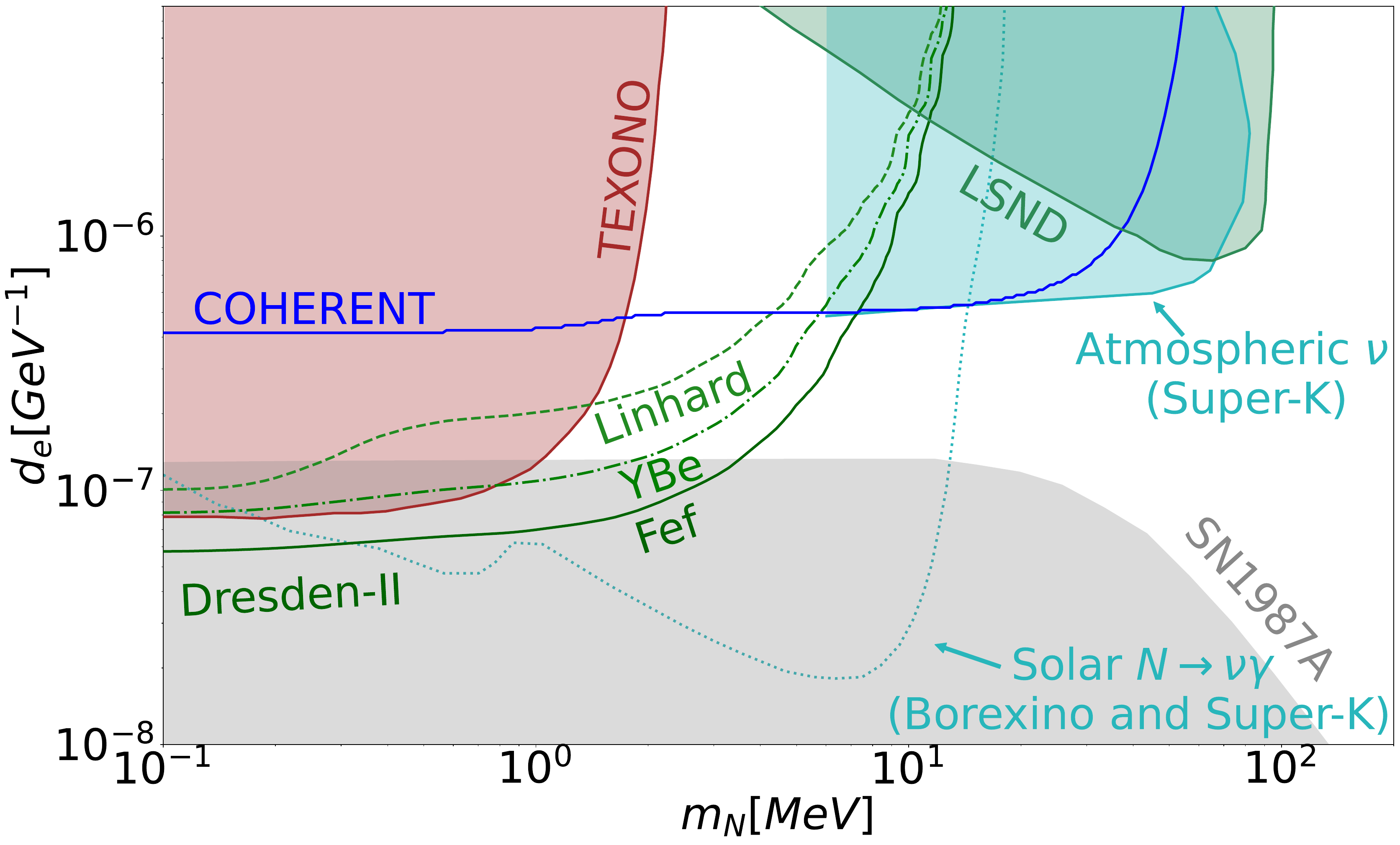}
		\includegraphics[scale=0.16]{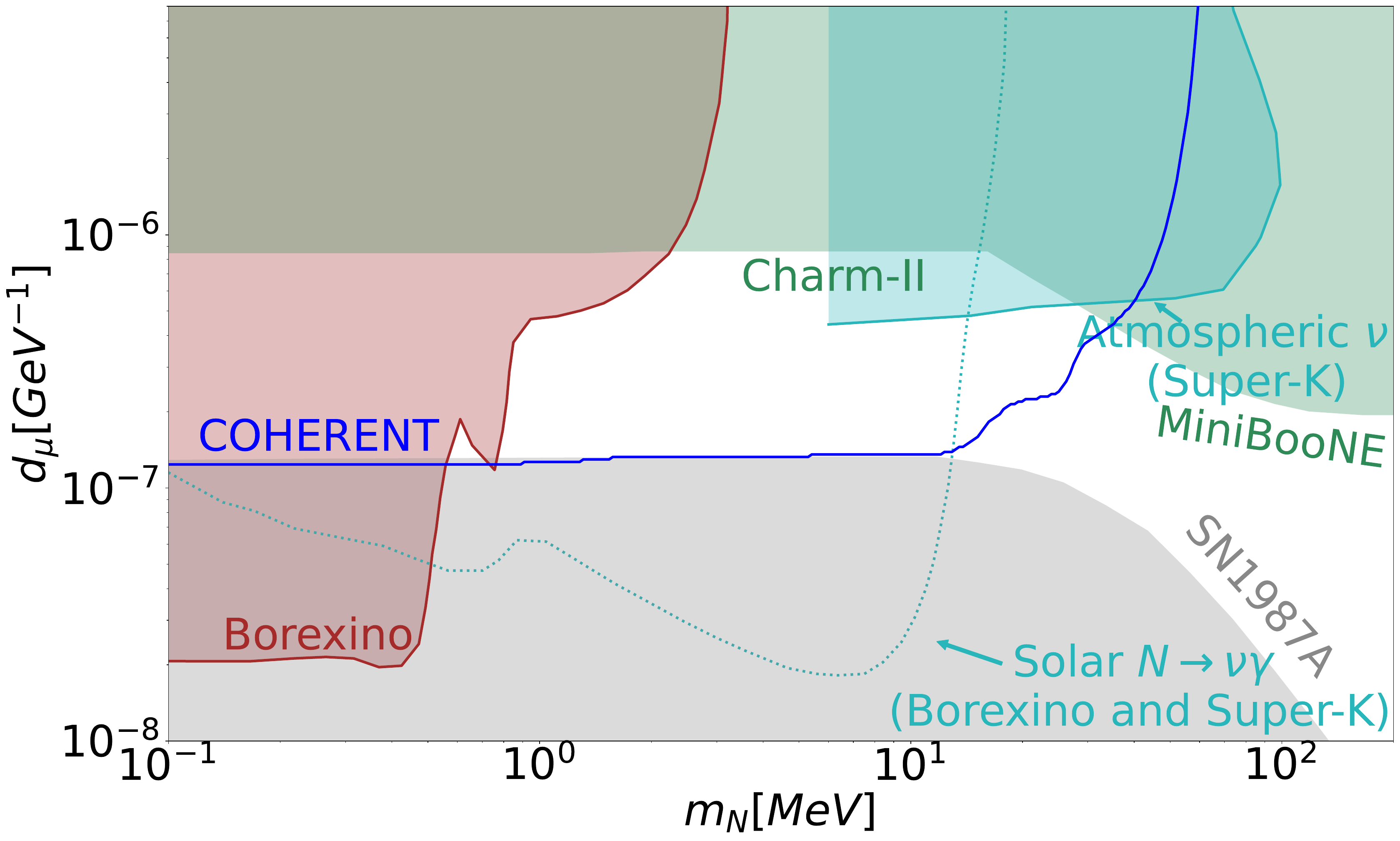}
        \includegraphics[scale=0.16]{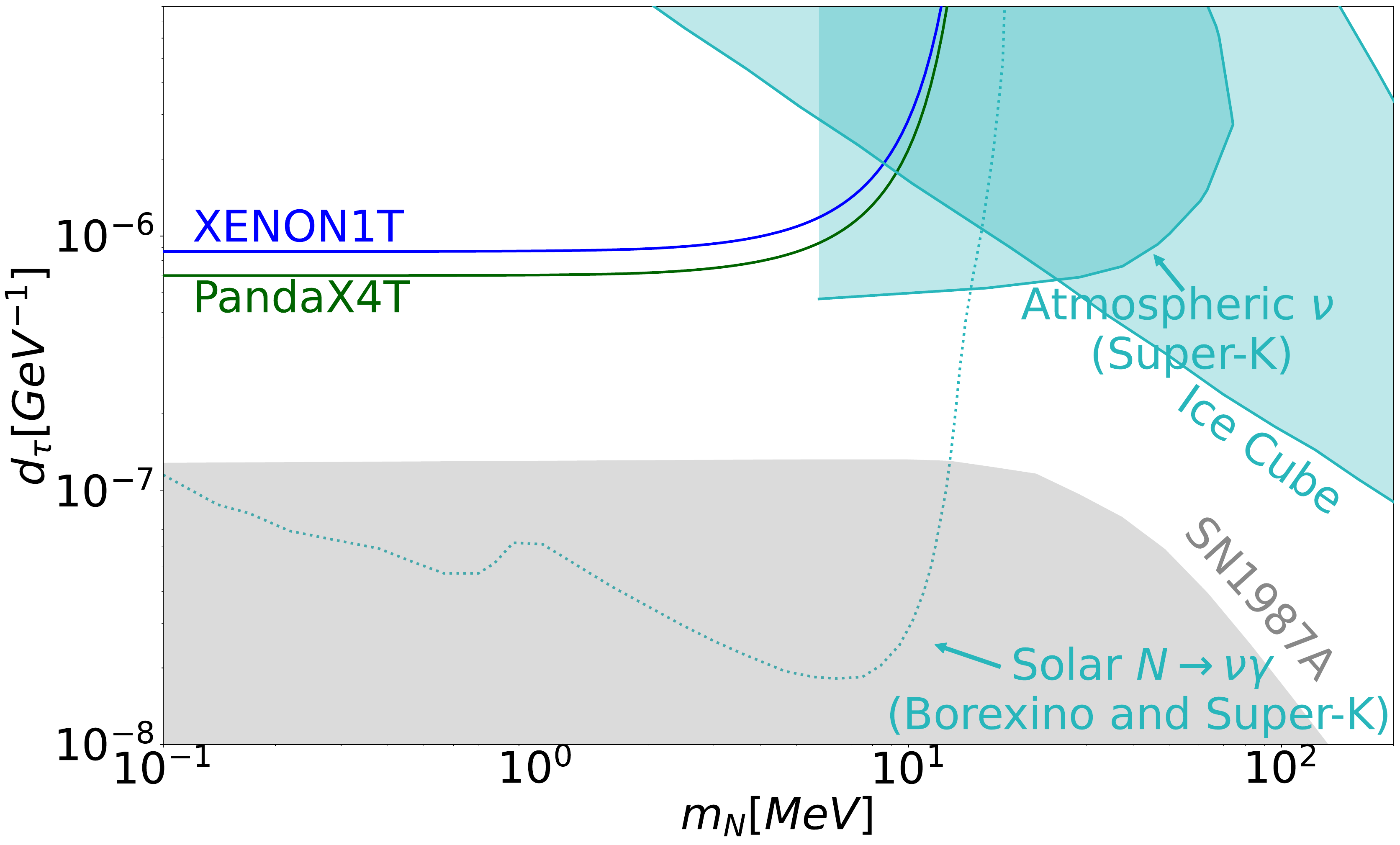}
		\caption{Exclusion limits for the $d_{\ell}$ of a specific flavor at 90$\%$ C.L. From top to bottom, results are shown for
		$d_{e}$, $d_{\mu}$, $d_{\tau}$. Figure adapted from Ref.\cite{Li:2024gbw}, see the text for more details.}
		\label{fig:sens}
\end{figure}

In Figure~\ref{fig:sens}, which is adapted from Ref.\cite{Li:2024gbw}, we illustrate the exclusion limits for the $d_{\ell}$ of a specific flavor at 90$\%$ C.L.
From top to bottom, the results are shown for $d_{e}$, $d_{\mu}$, $d_{\tau}$.
For the \cevns COHERENT analysis, the bounds obtained using the energy and time information are shown in blue solid lines.
For the Dresden-II reactor antineutrino analysis, the bounds are shown with green solid, dash-dotted and dashed lines using three different quenching models.
The limits on $d_{\tau}$ from the solar neutrino \cevns experiments
XENON1T~\cite{XENON:2020gfr} and PandaX-4T~\cite{PandaX:2022aac}
are shown with blue and green solid lines, respectively.
Constraints from other experiments are also shown for comparison, including SN1987A~\cite{Magill:2018jla}, TEXONO~\cite{TEXONO:2009knm}, LSND~\cite{Magill:2018jla}, atmospheric $\nu$ measurement at Super-Kamiokande~\cite{Gustafson:2022rsz}, looking for atmospheric neutrino upscattering to $N$ with subsequent decays, Borexino solar neutrino elastic scattering~\cite{Brdar:2020quo,Redchuk:2020hjv}, solar neutrino induced $N\to \nu\gamma$ measurements from Borexino and Super-Kamiokande experiments~\cite{Plestid:2020vqf}, MiniBooNE~\cite{Shoemaker:2018vii,Magill:2018jla,MiniBooNE:2007uho}, Charm-II~\cite{Shoemaker:2018vii,CHARM-II:1989srx} and IceCube~\cite{Coloma:2017ppo}.
The results described above concern flavor-specific dipole portals, except for the constraints from SN1987A, where flavor-universal couplings are assumed.

For $d_e$, the intense $\bar\nu_{e}$ flux of the Dresden-II reactor provides stringent constraints in the low mass range of Figure~\ref{fig:sens},
surpassing the limits from Borexino and Super-Kamiokande, which searched for solar neutrino upscattering to $N$ with subsequent decays~\cite{Gustafson:2022rsz}, as well as those from TEXONO~\cite{TEXONO:2009knm}, which measured reactor $\bar\nu_{e}$ \eves.
The constraints from COHERENT, based on a fit that includes both the energy and timing information, extend to $m_N \sim 50\,\text{MeV}$. The COHERENT bound is as competitive as those from LSND~\cite{Magill:2018jla} and atmospheric 
$\nu$ measurements at Super-Kamiokande, which are based on the non-observation of atmospheric neutrino upscattering to $N$ with subsequent decays~\cite{Gustafson:2022rsz}.
It is important to note that different choices of quenching models in the analysis of the Dresden-II \cevns data can significantly affect the constraints in the low mass region.

The COHERENT \cevns data provide the most stringent constraints on $d_\mu$ in the energy range at tens of MeV, with limits reaching as low as about $1\times10^{-7} \, \text{GeV}^{-1}$ below 10 MeV.
By using both the energy and time information of the COHERENT \cevns data,
the effects induced by the $\nu_\mu$ flux can be distinguished from those induced by the $\bar{\nu}_\mu$ and $\bar{\nu}_e$ fluxes
and different limits are obtained for $d_e$ and $d_\mu$.
For comparison, the constraints from other experiments such as MiniBooNE~\cite{Shoemaker:2018vii,Magill:2018jla,MiniBooNE:2007uho}, CHARM-II~\cite{Shoemaker:2018vii,CHARM-II:1989srx},
solar neutrino \eves observations at Borexino~\cite{Brdar:2020quo,Redchuk:2020hjv},
and atmospheric $\nu$ measurements at Super-K~\cite{Gustafson:2022rsz} are also illustrated in Figure~\ref{fig:sens}.

Since a substantial portion of the solar neutrino flux detected on Earth is made of $\nu_{\tau}$,
the \cevns measurements of solar neutrinos in dark matter direct detection experiments offer an opportunity
to probe the value of $d_{\tau}$.
The limits on $d_{\tau}$ from XENON1T~\cite{XENON:2020gfr} and PandaX-4T~\cite{PandaX:2022aac}
in Figure~\ref{fig:sens}
have been obtained from the 90\% C.L.
upper limits on the $^8$B solar neutrino flux established by these two experiments with \cevns data.
The XENON1T and PandaX-4T upper limits on $d_{\tau}$ are below $10^{-6}\,\text{GeV}^{-1}$
for $m_N \lesssim 10 \, \text{MeV}$
and agree with the more stringent Borexino and Super-Kamiokande limit.
Taking into account the recent measurements of \cevns produced by $^8$B solar neutrinos by the
PandaX-4T~\cite{PandaX:2024muv} and XENONnT~\cite{XENON:2024ijk} experiments,
we expect substantial improvements in the search of neutrino magnetic moments with solar \cevns
experiments in the future.

\section{Conclusions and future prospects}
\label{sec:future}

The search of neutrino electromagnetic interactions
is an important tool for testing the Standard Model (SM)
and for discovering new physics Beyond the Standard Model (BSM).
The SM can be tested by measuring the charge radii
of the three flavor neutrinos predicted by the calculation of
the SM radiative corrections.
The observation of different values of the neutrino charge radii
or other neutrino electromagnetic interactions
would be a discovery of BSM physics.

The current limits in Table~\ref{tab:CR}
on the values of the flavor neutrinos charge radii
are only about one order of magnitude larger than the
SM predictions in Eq.\ref{eq:SMchr}.
Therefore, the neutrino charge radii
may be the first neutrino electromagnetic property
that will be discovered in future experiments with higher precision.
However,
most experimental and theoretical studies
are devoted to neutrino magnetic moments,
which are predicted by many BSM models
and are connected with the neutrino masses.
The current upper bounds on the neutrino magnetic moments
in Tables~\ref{tab:MM-SBL} and~\ref{tab:MM-AST},
obtained from the data of laboratory experiments
and analyses of astrophysical data,
are more than seven orders of magnitude larger
than the predictions of the simplest extensions of the SM
(see Eqs.\ref{muD}--\ref{muDtranbf}).
However,
the search of the effects of the neutrino magnetic moments
below the current limits
is motivated by the predictions of
larger values of the neutrino magnetic moments
by more elaborate BSM models
with Dirac~\cite{Bolanos-Carrera:2023ppu}
or Majorana~\cite{Pal:1981rm,Barr:1990um,Babu:1990wv,Pal:1991qr,Boyarkin:2014oza,Lindner:2017uvt}
neutrinos.
The neutrino charges, sometimes called ``millicharges'',
are usually considered more exotic,
but they are worth searching for,
since they are associated with a possible
charge dequantization~\cite{Foot:1990uf,Foot:1992ui,Giunti:2014ixa,Das:2020egb}.

The current best limit on a neutrino magnetic moment
obtained in a laboratory experiment
is the bound on $\mu_{\nu_{e}}$ obtained
by the GEMMA~\cite{Beda:2012zz} experiment
(see Table~\ref{tab:MM-SBL}).
Most of the astrophysical limits on the neutrino magnetic moments
in Table~\ref{tab:MM-AST}
are more stringent,
but they suffer from astrophysical uncertainties.
Therefore,
it is important to continue the search of the effects
of the neutrino electromagnetic properties in laboratory experiments
where the neutrino beam is under control.
Note that novel detection methods such as the enhancement of the screening effect in semiconductor detectors~\cite{Li:2022pxj} could push the limit on the neutrino magnetic moment down to $10^{-13}\,\mu_{\text{B}}$.

Future improvements in the search of neutrino electromagnetic interactions
are expected from the results of the
GEMMA~II~\cite{Alekseev:2017yla},
TEXONO~\cite{Singh:2017bns},
JUNO-TAO~\cite{JUNO:2020ijm,Brdar:2024lud},
CLOUD~\cite{Brdar:2024lud},
SHIP~\cite{SHiP:2021nfo},
FPF@LHC~\cite{MammenAbraham:2023psg},
SBN~\cite{Machado:2019oxb},
ESS~\cite{Baxter:2019mcx},
DUNE~\cite{Mathur:2021trm}
experiments,
\cevns experiments~\cite{Barbeau:2021exu,Abdullah:2022zue},
\cevas\footnote{Coherent elastic neutrino-atom scattering.} experiments~\cite{Cadeddu:2019qmv}, in particular the SATURNE project~\cite{SATURNE:2024vmu},
experiments with a radioactive~\cite{Herrera:2024ysj}
or an IsoDAR~\cite{IsoDAR:2024rvi} source
near an appropriate detector,
as well as from analyses of improved astrophysical data,
especially the dark matter and solar neutrino experiments
PandaX~\cite{PandaX:2024muv},
XENONnT~\cite{XENON:2024ijk,Herrera:2023xun},
LUX-ZEPLIN~\cite{LZ:2021xov,LZ:2023poo},
XMASS~\cite{XMASS:2020zke},
DARWIN~\cite{DARWIN:2020bnc,Giunti:2023yha},
XLZD~\cite{XLZD:2024gxx},
DarkSide~\cite{DarkSide-20k:2024yfq}.

In conclusion, we expect an interesting future for the search of neutrino electromagnetic properties.

\section*{DISCLOSURE STATEMENT}
The authors are not aware of any affiliations, memberships, funding, or financial holdings that
might be perceived as affecting the objectivity of this review. 

\section*{ACKNOWLEDGMENTS}
We would like to thank
Garv Chauhan,
Andreas Lund,
and
Jack Shergold
for useful comments on the first version of this review.
The work of C. Giunti is partially supported by the PRIN 2022 research grant Number 2022F2843L funded by MIUR.
The work of K. Kouzakov and A. Studenikin is supported by the Russian Science Foundation (project No. 24-12-00084).
The work of Y.F. Li is partially supported  by National Natural Science Foundation of China under Grant Nos.~12075255 and 11835013.

\bibliographystyle{ar-style5}
\bibliography{main}

\end{document}